# Spinning Driven Dynamic Nuclear Polarization with Optical Pumping

*Frederic Mentink-Vigier[a]\*, Vinayak Rane[b,c]\*, Thierry Dubroca[a], Krishnendu Kundu[a]*

[a]National High Magnetic Field Laboratory, Florida State University, 1800 E Paul Dirac Drive, Tallahassee, FL, 32310, USA

[b]Bhabha Atomic Research Centre, Trombay, Mumbai 400085, India

[c]current address: Indian Institute Of Geomagnetism, Plot 5, Sector 18, New Panvel (W), Navi Mumbai - 410218, Maharashtra, India.

fmentink@magnet.fsu.edu

rane.vinayak@gmail.com

We propose a new, more efficient, and potentially cost effective, solid-state nuclear spin hyperpolarization method combining the Cross Effect mechanism and electron spin optical hyperpolarization in rotating solids. We first demonstrate optical hyperpolarization in the solid state at low temperature and low field, and then investigate its field dependence to obtain the optimal condition for high-field electron spin hyperpolarization. The results are then incorporated into advanced Magic Angle Spinning Dynamic Nuclear Polarization (MAS-DNP) numerical simulations that show that optically pumped MAS-DNP could yield breakthrough enhancements at very high magnetic fields. Based on these investigations, enhancements greater than the ratio of electron to nucleus magnetic moments (>658 for 1H) are possible without microwave irradiation. This could solve at once the MAS-DNP performance decrease with increasing field and the high cost of MAS-DNP instruments at very high fields.

**Introduction**

MAS-DNP is a powerful solid-state NMR (ssNMR) method that reduces the duration of ssNMR experiments by orders of magnitude.[1] In short, the high polarization of paramagnetic species stemming from microwave (μw) irradiation at the Larmor frequency, can be transferred to nuclei to enable molecular-level characterization even when the isotope of interest is in low concentration or has low receptivity.[2–5] Over the past two decades, there has been significant progress in the development of hardware[6–11], sample preparation methods,[12–16] paramagnetic species used as sources for DNP[17–22] and the theoretical understanding of MAS-DNP.[23–26] MAS-DNP most commonly uses biradicals[27] to generate the nuclear hyperpolarization via the Cross Effect (CE) mechanism, which involves fast energy level anti-crossing.[23–26,28,29]

As in conventional ssNMR, very high field MAS-DNP (>14.1 T/600 MHz) enables higher resolution, but faces multiple challenges, such as reduced efficiency of the CE with the field[30,31] and significant microwave absorption at high frequencies,[11] which reduce the large electron polarization difference and the concomitant nuclear polarization enhancement.[25,30,32,33] Finally, the significant cost of high-field MAS-DNP instrumentation limits widespread availability.

In parallel to DNP developments, optical irradiation has been used to improve NMR sensitivity. For example, nuclear hyperpolarization in ssNMR experiments via photo-CIDNP was observed in certain systems.[34–40] Furthermore, optical electron spin hyperpolarization offers a promising approach for carrying out liquid-state DNP[41,42] and hyperpolarized triplet state has been combined with Integrated Solid-Effect at low field to generate nuclear spin hyperpolarization.[43–45]

In this letter, we propose a novel method that can provide much higher hyperpolarization than traditional MAS-DNP at high fields, in addition to addressing the issues listed above via the use of optical electron spin hyperpolarization in the solid state. The concept, dubbed optically pumped

MAS-DNP (MAS-OPDNP), uses optical irradiation to photophysically generate the electron spin polarization difference required for the CE mechanism and build on the effect of the sample's rotation to hyperpolarize the nuclei. This concept enables nuclear spin hyperpolarization that is not restricted to the ratio of electron to nucleus magnetic moments, while potentially using affordable hardware. Finally, the method is expected to be field independent, and therefore, it should be easily added to most modern ssNMR spectrometers.

The present work first demonstrates experimentally, in the solid state and X-band (low field), that optically driven electron spin hyperpolarization is possible for nitroxides commonly used for MAS-DNP.[27,46,47] We subsequently investigate the field dependence of optical electron spin hyperpolarization and its characteristic time scales. Based on these results, we propose a Chromophore-Radical-Radical Polarizing Agent and present simulations that were conducted with a high-performance MAS-DNP numerical tool,[25,48] to explore the potential of the CE MAS-DNP mechanism at high magnetic field using optical electron spin hyperpolarization.

**Results**

*Photophysical electron spin hyperpolarization* - Optically pumped electron spin hyperpolarization can be generated in chromophore-radical (CR) systems during photophysical quenching processes.[49–57] The electron hyperpolarization generation in CR systems is well understood in the liquid state.[58–62] two mechanisms contribute to the electron hyperpolarization (see energy diagram in Scheme 1): (i) the Spin-Orbit induced Inter-System-Crossing (SO-ISC)[62,63] and (ii) the $D_1$-$Q_1$ conversion via the Reverse Quartet Mechanism (RQM).[64] The first mechanism is due to the spin-orbit coupling, while the second involves the large Zero-Field Splitting (ZFS), $D_{ZFS}$, both in the excited triplet state of the chromophore.

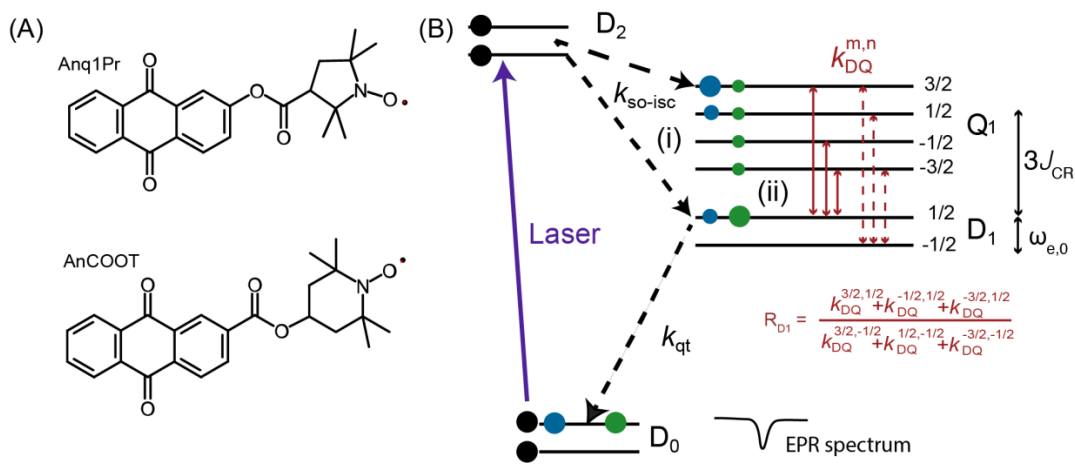

Scheme 1: (A) Structures and acronyms of the molecules examined in this study. (B) Key processes involved in electron spin hyperpolarization in a Chromophore-Radical adduct at low magnetic field (~0.3 T). The laser pulse excites the molecule from the $D_0$ to the $D_2$ state (termed sing doublet),[56,65] which then decays from $D_2$ to $Q_1/D_1$ (mixing of triplet and doublet) via SO-ISC, followed by the quenching of $D_1$ state (termed trip-doublet) to $D_0$ at a rate of $k_{qt}$. Two mechanisms lead to hyperpolarization in the $D_0$ state: (i) the selective transition of the SO-ISC and (ii) the reversible transitions between $D_1$ and $Q_1$ states (RQM). Black dots represent the initial photoexcited populations, the blue and green dots denote the population redistributions after process (i) and (ii), respectively. For process (ii), after decay via SO-ISC, the populations in the $Q_1$ state are equilibrated via longitudinal relaxation and driven by the RQM pathway via the $D_1 \rightarrow Q_1$ transitions thanks to cross-relaxation, with rates defined by the $k_{DQ}^{m,n}$ between $Q_1^m$ and $D_1^n$ (red arrows) to generate hyperpolarization in the $D_0$ state. The ratio of rates $R_{D1} = \Sigma k_{DQ}^{m,1/2}/\Sigma k_{DQ}^{m,-1/2}$ (red solid and red dashed arrows) quantifies the selective polarization in the $D_1$ states. $\omega_{e,0}$ represents the Zeeman Larmor frequency and $3J_{CR}$ ($< 0$) represents the magnitude of the exchange splitting between the $Q_1$ and $D_1$ states.

Recently, ANCOOT and Anq1PR (Scheme 1A), two efficient CR-systems, were reported to generate a large electron hyperpolarization in solution.[60,66] We examined their hyperpolarization efficiency in the solid state as a preliminary assessment for MAS-DNP applications. Fig. 1A shows the EPR spectra of ANCOOT in toluene at 100 K in thermal equilibrium (black curve) and hyperpolarized (blue curve) states. The hyperpolarized EPR spectrum is emissive and a near mirror image of the thermal equilibrium one, without any signature of the quartet EPR spectrum. From the signal intensity ratio, we estimated the electron polarization enhancement to be about -100 times the thermal spin polarization (~ -30 % polarization, see SI for details on the evaluation of experimental hyperpolarization). The hyperpolarization is generated on a very fast time-scale < 100 ns (the instrument response time is 100 ns), which is important for applications in MAS-DNP.

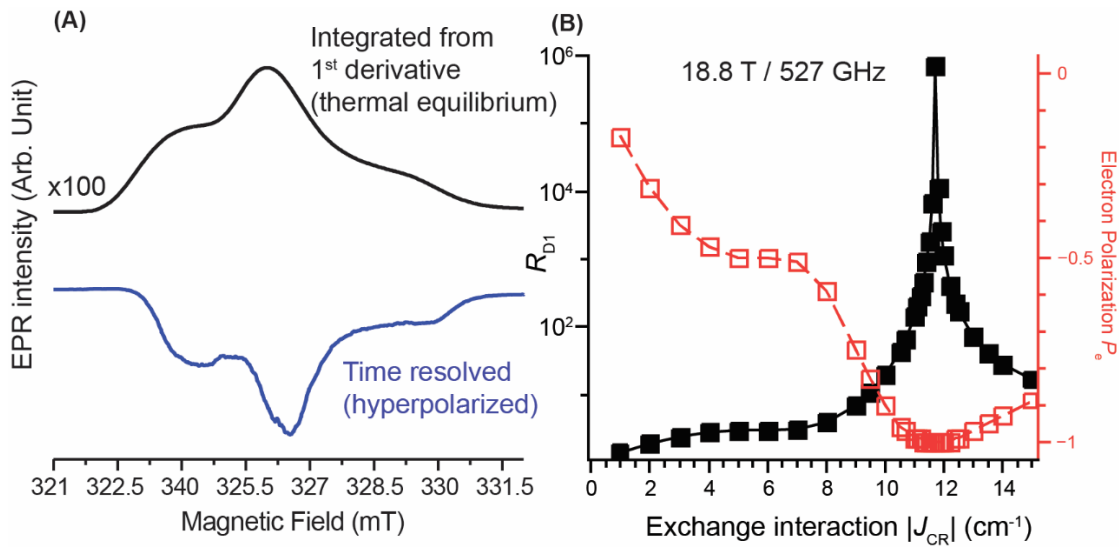

Fig. 1: (A) Experimental EPR spectrum (integrated from the first derivative) of ANCOOT in toluene at 100K (black line) and its hyperpolarized EPR spectrum recorded for 2.5 μs after a 355 nm laser pulse (blue line) and integrated using a boxcar averager over 0.3 μs. See SI for experimental details. The gain factor of -100 for the Boltzmann signal represents the ESP enhancement obtained from time resolved experiments (TREPR), see appendix 4. (B) Plot of the RQM selectivity factor, $R_{D1}$ (black squares), and the corresponding values of $P_e$ (red squares) as a function of the exchange parameter, $|J_{CR}|$ ($J_{CR}<0$), for the case of pure RQM, i.e., process (ii) (scheme 1). N.B. solid black and dashed red interpolating lines are provided as guides to the eye.

The solid state hyperpolarized EPR spectrum and its time dependence for the anthraquinione-nitroxides are determined by the SO-ISC and RQM mechanisms. Their relative contributions at X-band frequencies are beyond the scope of the current work and will be described in a future publication.

At high magnetic fields the SO-ISC mechanism may become less efficient as the net (rotationally invariant) component of the polarization generated during the ISC has an inverse field dependence (see eq. S10).[52,67] It is therefore suspected that the SO-ISC induced hyperpolarization would likely decrease by orders of magnitude at high magnetic field (18.8 T). Thus, at high field and low temperature (~100 K), the RQM is likely the only mechanism that can generate the electron hyperpolarization needed to observed OPDNP.

The polarization generated by the RQM is determined by the rates $k_{DQ}^{m,n}$ (see Eq. (1), Scheme 1 and Appendix 2, eq. (S11)) which depends on the ZFS and exchange interaction ($J_{CR}$) between chromophore and radical in the excited $D_1$-$Q_1$ state.[59,68] Due to the solid state nature of the sample, $k_{DQ}^{m,n}$ would also depend on the orientation of the ZFS tensor with respect to the magnetic field. However, the ZFS (~0.3 cm$^{-1}$) is relatively small compared to the electron Zeeman term at high fields (>5 cm$^{-1}$), and the exchange interaction. Thus, the manifestation of anisotropy in $k_{dq}$ is expected to be weak. Furthermore, since the CR is dissolved in a glass matrix, all crystal orientations are present in the matrix, we thus used average RQM rates, as it is done in the liquid state:[64]

$$k_{DQ}^{mn} \propto \frac{\overline{\langle Q_1^m | H_{ZFS} | D_1^n \rangle^2}}{\Delta E(Q_1^m, D_1^n)^2} \qquad \text{Eq. (1)}$$

where $m \in \left[-\frac{3}{2}, -\frac{1}{2}, \frac{1}{2}, \frac{3}{2}\right]$, $n \in \left[-\frac{1}{2}, \frac{1}{2}\right]$, and $\Delta E(Q_1^m, D_1^n)$ is the energy difference between states $Q_1^m$ and $D_1^n$. An estimate of $k_{DQ}^{mn}$ at a low temperature was obtained by numerical fitting of the time resolved EPR time profile of ANCOOT recorded at 100 K (see Eq. S7).

At high fields, $k_{DQ}^{mn}$ can be tuned by adjusting $\Delta E(Q_1^m, D_1^n) \approx 3J_{CR} + (m-n)\omega_{e,0}$, which is dominated by $J_{CR}$ and the Larmor frequency of the electron $\omega_{e,0}$. The optimal value of $J_{CR}$ to maximize the efficiency of the RQM can be determined through the selectivity factor, $R_{D1}$, which is the ratio of the sum of all the rates from $Q_1$ levels to the $D_1^{1/2}$ and the $D_1^{-1/2}$ energy levels, written as:

$$R_{D1} = \frac{k_{DQ}^{Q_1^{+3/2} D_1^{+1/2}} + k_{DQ}^{Q_1^{-1/2} D_1^{+1/2}} + k_{DQ}^{Q_1^{-3/2} D_1^{+1/2}}}{k_{DQ}^{Q_1^{+3/2} D_1^{-1/2}} + k_{DQ}^{Q_1^{+1/2} D_1^{-1/2}} + k_{DQ}^{Q_1^{-3/2} D_1^{-1/2}}} \qquad \text{Eq (2)}$$

$R_{D1}$ quantifies how the spin populations redistribute itself within the $D_1$ states, which ultimately dictate the nitroxide hyperpolarization (see Appendix 3 in the SI). $R_{D1}$ was calculated at 18.8 T (a typical high field for MAS-DNP) and the resulting plot (Fig. 1B) reveals a large RQM efficiency for $|J_{CR}|$ in the range of 9 - 14 cm$^{-1}$. It is maximum at ~ 12 cm$^{-1}$ where $\Delta E(Q_1^{-3/2}, D_1^{1/2}) = 0$. In this case, the $D_1^{1/2} \leftrightarrow Q_1^{-3/2}$ transition becomes the dominant RQM pathway, as the mixing rate constants originating from the other $D_1$-$Q_1$ transitions are too small to have an effect (see figure

S3). This special case enables a selective enhancement of the population of $D_1^{1/2}$ state and the generation of a very large electron hyperpolarization in the $D_0$ state via the $D_1 \rightarrow D_0$ pathway.

The model predicts an electron hyperpolarization level after laser irradiation, $P_e$ (red curve, Fig.1B, see SI for derivation), that can reach $-1$ at the optimal $J_{CR}$; however, it also shows that smaller exchange interactions (i.e., 4-8 cm$^{-1}$) already yield significant hyperpolarization $P_e \approx -0.5$. Such exchange interactions can be attained in existing CR systems given that earlier studies on the chromophore TEMPO showed $J_{CR}$ values in the range of 1-5 cm$^{-1}$.[64]

Hence, we conclude that solid state optical electron spin hyperpolarization is possible at X-band frequencies (Fig. 1). In addition, at high fields, a strong electron hyperpolarization in the solid state can be obtained via an "RQM-only" mechanism, provided that the $J_{CR}$ falls within a favorable range. In turn, this allows us to explore the potential of MAS-DNP using optically pumped hyperpolarized nitroxides.

*Optically Pumped Cross-Effect for MAS-DNP.* From the mechanistic analysis of the photophysical hyperpolarization, it is now possible to assess how optical pumping could benefit CE under MAS-DNP. CE MAS-DNP requires the use of biradicals, i.e., molecules with two coupled unpaired electrons in their ground state.

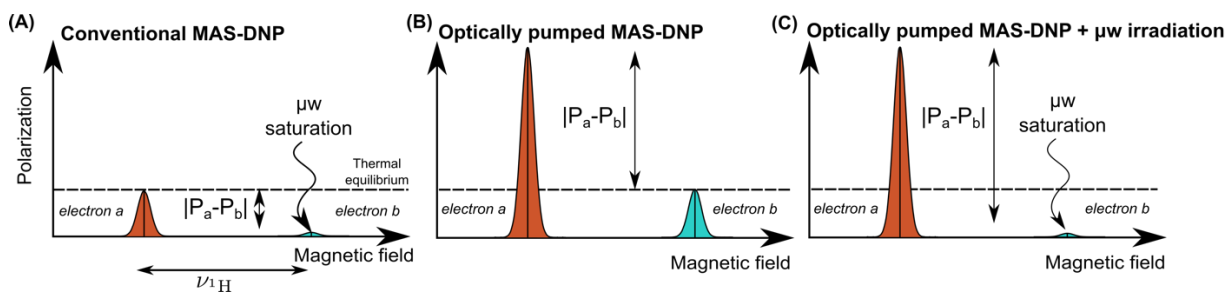

Scheme 2: Schematics of the proton CE MAS-DNP mechanism in the ideal case of a biradical with two non-overlapping EPR spectra separated by the proton Larmor frequency. (A) Conventional MAS-DNP where the spins of electron *b* are saturated, while the spins of the electron *a* are at thermal equilibrium. The proposed optically pumped hyperpolarization method in which the polarization of electron spin *a* is increased while the electron *b* is at thermal equilibrium (B) or saturated by microwave irradiation (C).

Scheme 2 shows an ideal CE biradical model with two interacting moieties (a) and (b) with electron Larmor frequencies, $v_{a(b)}$, and difference matching the Larmor frequency of the proximate nuclear spins,[33,69,70] $v_n$, such that $|v_a - v_b| \sim |v_n|$. When an electron spin polarization difference, $|P_{e,a} - P_{e,b}|$, is generated, this results in nuclear hyperpolarization.[32,33] Under DNP, it is the μw irradiation that generates $|P_{e,a} - P_{e,b}|$ (Scheme 2A). The nuclear polarization, $|P_n|$, in a steady state, is related to $|P_{e,a} - P_{e,b}|$ by[32]

$$|P_{e,a} - P_{e,b}| \geq |P_n| \qquad \text{Eq. (3)}$$

If instead *a* is hyperpolarized via optical means, two other cases can be envisioned: (i) *a* is hyperpolarized, and *b* is at thermal equilibrium (Scheme 2B); or (ii) *a* is hyperpolarized, and *b* is saturated with a microwave irradiation (Scheme 2C).

To realize the concept presented in Scheme 2B and 2C, we need a molecule comprised of a chromophore and biradical CR$_a$-R$_b$ where C and R$_a$ are much closer to each other than C and R$_b$, such that in the *excited state* $|J_{CRa}| \sim$ 5-14 cm$^{-1}$ and $J_{CRb} \sim 0$ cm$^{-1}$. In addition, in the ground state, the biradical, R$_a$-R$_b$, should have similar properties to typical biradicals that are used as MAS-DNP polarizing agents. For example, biradicals comprised of Trityl and nitroxide moieties are known to be efficient for MAS-DNP[21,69,71]. The three cases in Scheme 2 were simulated using a fictitious "CR$_a$-R$_b$ = Chromophore-TEMPO-Trityl" molecule represented at the top of Fig. 2 (herein, electron "*a*" stands for the nitroxide and "*b*" for the Trityl, unless otherwise specified).

Under MAS, this spin system undergoes fast energy level anti-crossings (or rotor-events)[32] because the EPR spectra of Trityl and TEMPO are anisotropic, and therefore overlap with one another. This means that the nuclear hyperpolarization results from CE rotor events that transfer the polarization from the electron pair to the nuclei. These rotor-events are active due to MAS, and it is important to note that they can perturb the nuclear spin polarization even in absence of µw irradiation.[72,73] In addition, because the EPR spectra overlap, the dipolar/exchange rotor events are active. This type of rotor event is key for maintaining the electron polarization difference and ensures that the transfer of polarization to the nuclei has a constant sign, allowing for large polarization buildups.[32,33]

The complexity of this mechanism and its dependence on relaxation properties requires treatment with numerical simulations. Therefore, we used the "Box model", which accounts for multiple three-spin systems {2 electron spins – 1 proton spin} distributed in a bounded space that has been extensively tested and validated.[25,48,74] This model treats the inter-biradical interactions to mimic those of a 10 mM biradical solution (see SI), thus faithfully represents the spin dynamics of the electrons[25]. The model is modified to account for the optical hyperpolarization by assuming that under continuous (or pulsed) laser irradiation nitroxide hyperpolarization is generated on a time scale faster than the MAS period, as determined by the RQM analysis (see SI for details).

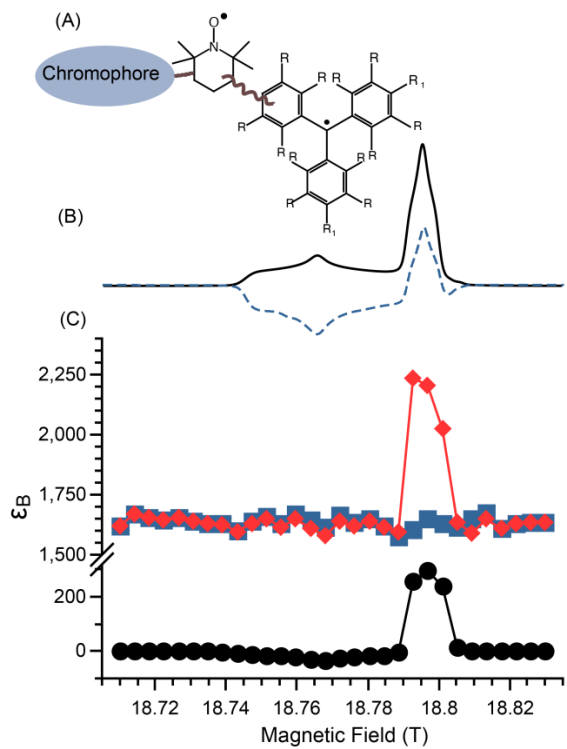

Fig. 2: (A) Fictitious "Chromophore-TEMPO-Trityl" $CR_a$-$R_b$ molecule. The wavy line represents unspecified bonds. (B) Simulated EPR spectra of $CR_a$-$R_b$ in thermal equilibrium (solid black line) with laser irradiation assuming $P_{e,a} = -0.25$ (dashed blue line). (C) simulations of the MAS-DNP field profiles using the modified Box model for conventional MAS-DNP (black dots), optically hyperpolarized MAS-DNP (blue squares), and combined μw and optical irradiated hyperpolarization (red diamonds). In all simulations, the optical hyperpolarization leads to $P_{e,a} \to -0.75$, and for clarity we plot $\epsilon_B = -f(B_0)$. Details about the spin system are given in the SI.

Fig. 2C displays the nuclear spin polarization gain as function of the magnetic field (see SI for calculation details) for conventional MAS-DNP (black dots), optical hyperpolarization (blue squares), and optical hyperpolarization combined with μw irradiation (red diamonds). The field profile calculated for conventional MAS-DNP spans the entire EPR spectra of the Trityl-TEMPO

(shown in Fig. 2B) and has a sharp feature at the Trityl Larmor resonant frequency [21,69,71]. In this case, the maximum polarization gain is $\epsilon_B \approx 295$ with the chosen simulation parameters (see SI). On the other hand, the field profile in the presence of both optical and μw irradiation is very similar in shape but present now a staggering maximum value, $|\epsilon_B| \approx 2200$.

While this result is outstanding, the curve with blue squares, which reports $\epsilon_B$ in the case of optical hyperpolarization (only) is also very encouraging: it predicts $|\epsilon_B| \approx 1700$. This is seven times higher than in the conventional MAS-DNP case and also corresponds to the baseline of the optical and μw irradiation cases. This enhancement is the result of the CE mechanism being always active under MAS.[23,33] In absence of μw irradiation, this can gives rise to nuclear depolarization[72,73] for bis-nitroxides as $|P_{e,a} - P_{e,b}|_{\mu w,off} \leq |P_n|_{eq}$, while Trityl-nitroxides do not depolarize significantly, i.e., $|P_{e,a} - P_{e,b}|_{\mu w,off} \approx |P_n|_{eq}$. The centers of mass for the Trityl and nitroxide EPR spectra are separated by the proton Larmor frequency, leading to little depolarization ($\epsilon_B \approx 1$), as seen outside of the EPR resonant field (Fig. 2C, black dots)[21]. For the CR$_a$-R$_b$, this separation of the centers of mass of the EPR spectra is key for efficient DNP with optical hyperpolarization, enabling the existence of an electron spin polarization difference. At $B_0 = 18.8$ T and 100 K, the thermal equilibrium polarization of the Trityl is $P_{e,b}^{eq} \approx 0.12$, thus with electron hyperpolarization of the nitroxide (Fig. 2, blue square and red diamonds) we have:

$$|P_{e,a} - P_{e,b}^{eq}| \gg |P_n|_{eq} \qquad \text{Eq. (4)}$$

which explains the large $|\epsilon_B|$.

A broad range of scenarios were explored, which report $\epsilon_B$ as a function of the nitroxide hyperpolarization level with and without μw (Fig. 3A). Both sets of simulations display linear trends, with a steeper slope when biradicals are under μw irradiation (due to the larger

$|P_{e,a} - P_{e,b}|$). Enhancements $|\epsilon_B|$ larger than the ratio of electron to proton magnetic moments (~658) can be achieved for $P_{e,a} \to 0$ or $P_{e,a} \to 2P_{e,b}^{eq} = 0.24$. However, with the chosen parameters, the nitroxide electron hyperpolarization must be lower than −0.3 or higher than 0.4, as shown Fig. 3A. These values are larger than the ideal case, because $|P_{e,a} - P_{e,b}|$ is affected by the inter-biradical interactions. Under MAS, they tend to equilibrate the polarization among all the Trityls and the nitroxides contained in the Box,[25,72] thereby affecting the average electron spin polarization difference. This effect is spin-system dependent; therefore, different slopes (Fig 3A) are obtained for different electron relaxation times, magnetic fields, radical concentrations and/or temperatures (see examples in SI). The enhancement can also be calculated as a function of the magnetic field. For a given $P_{e,a}$, using (see SI for full derivation):

$$\epsilon_B = \frac{B_{eff} - B_0}{B_0} \qquad \text{Eq. (5)}$$

where $B_{eff}$ is an effective magnetic field.

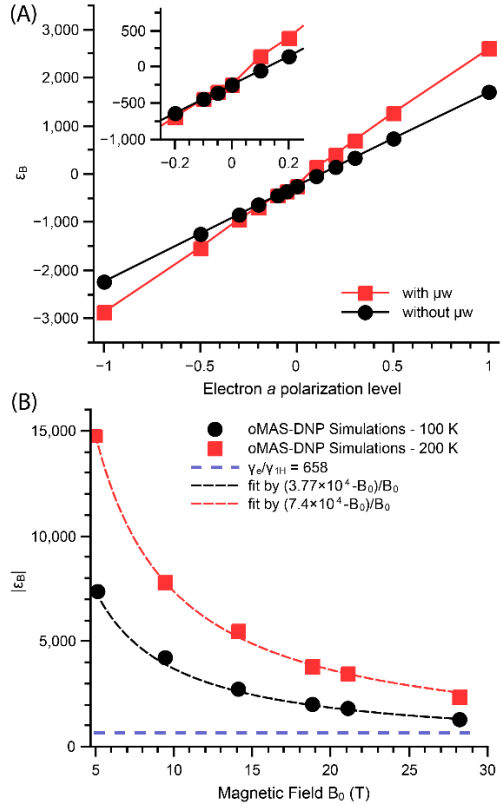

Fig. 3: (A) Calculated MAS-DNP polarization gain, $\epsilon_B$, for a Trityl-nitroxide biradical using the modified Box model, with μw irradiation (red squares), and without (black dots) as a function of the nitroxide hyperpolarization. (B) Calculated $|\epsilon_B|$ as function of magnetic field without μw irradiation, assuming a nitroxide hyperpolarization of $P_{e,a} \to -0.75$ at 100 K (black dots) and 200 K (red squares). All other simulation parameters were kept constant. Dashed lines: best fits using Eq. (5).

Fig. 3B displays $|\epsilon_B|$ as function of the field with $P_{e,a} \to -0.75$ and without μw irradiation, which confirms $|\epsilon_B| > 658$. The simulations carried out at 100 and 200 K perfectly fit with Eq. (5). At higher temperatures, the significantly larger $|\epsilon_B|$ is the result of the lower equilibrium polarizations for both Trityl electrons and the protons. This illustrates yet another potential benefits of MAS-OPDNP: better efficiency at higher temperature. Of course, this would depend on the relaxation times at higher temperature, where the CE mechanism may not be as efficient in terms of total nuclear polarization. Finally, the MAS-OPDNP simulations of a bis-nitroxide with a structure equivalent to "AMUPol"[48] are reported in the SI. While the EPR spectra of $R_a$ and $R_b$ have the same centers of mass, the resulting $|\epsilon_B|$ is on the order of 200 at 14.1 T, which is very similar to the conventional MAS-DNP for AMUPol[48].

**Conclusions**

In conclusion, the MAS-OPDNP simulations demonstrate that even moderate nitroxide hyperpolarization could easily lead to $|\epsilon_B| > 658$. This new concept might also work at higher temperatures, which may be beneficial for samples that require higher peak resolutions. The analysis of the hyperpolarization transfer mechanism reveals that the success of this proposed concept requires the synthesis of new $CR_a$-$R_b$ molecules possessing the appropriate exchange interactions in the excited state. We anticipate experimental challenges such as optical absorption and stability of the $CR_a$-$R_b$. However, this innovative approach removes the need for expensive high power μw sources and "sweepable" high-field NMR magnets. Instead, MAS-OPDNP would rely on much more affordable high-power lasers that are currently commercially available and sample spinning. Finally, the MAS-OPDNP concept has the potential to be a paradigm shift for high-field MAS-DNP, which will have broad impacts on characterization of chemical compounds, biological molecules, and numerous materials.

*Acknowledgments* Zhehong Gan and Robert Schurko are acknowledged for their critical assessment of the manuscript. The National High Magnetic Field laboratory (NHMFL) is funded by the National Science Foundation Division of Materials Research (DMR-1644779) and the State of Florida. A portion of this work was supported by the NIH P41 GM122698.*References*

# Supporting Information: Spinning Driven Dynamic Nuclear Polarization with Optical Pumping

*Frederic Mentink-Vigier[a]\*, Vinayak Rane[b,c]\*, Thierry Dubroca[a], Krishnendu Kundu[a]*

[a]National High Magnetic Field Laboratory, Florida State University, 1800 E Paul Dirac Drive, Tallahassee, FL, 32310, USA

[b]Bhabha Atomic Research Centre, Trombay, Mumbai 400085, India

[c]Current address: Indian Institute of Geomagnetism, Plot 5, Sector 18, New Panvel (W), Navi Mumbai - 410218, Maharashtra , India

[a]fmentink@magnet.fsu.edu,

[b]rane.vinayak@gmail.com

# Table of Contents

## Appendix 1: Hamiltonians used in the simulations

For CR in the excited $D_1$-$Q_1$ states the Hamiltonian is defined as:

$$\hat{H}_{CR}^{D1-Q1} = \hat{H}_Z + \hat{H}_{JCR} + \hat{H}_{HF,14N} + \hat{H}_{ZFS} \qquad \text{Eq. (S1)}$$

where $\hat{H}_Z$ is the Zeeman Hamiltonian:

$$\hat{H}_Z = g_C \beta_e B_0 \hat{S}_{z,c} + g_R \beta_e B_0 \hat{S}_{z,R} \qquad \text{Eq. (S2)}$$

with $g_C$ and $g_R$ as the $g$ factors of the chromophore and the radical respectively and $B_0$ the external magnetic field. $\hat{H}_{JCR}$ is the exchange interaction and the strongest interaction defined as:

$$\hat{H}_{JCR} = -2J_{CR}\left(\hat{S}_{z,C}\hat{S}_{z,R} + \frac{1}{2}(\hat{S}_C^+\hat{S}_R^- + \hat{S}_C^-\hat{S}_R^+)\right) \qquad \text{Eq. (S3)}$$

$\hat{H}_{HF,14N}$ is the nitroxide hyperfine coupling to the $^{14}$N nuclei and is:

$$\hat{H}_{HF,14N} = A_{14N}\hat{S}_{zR}\hat{I}_{z,14N} \qquad \text{Eq. (S4)}$$

Finally, $\hat{H}_{ZFS}$ is the Zero-Field Splitting (ZFS) interactions. It is responsible for mixing the $D_1$ and $Q_1$ states. Note this is a three electrons system (two electrons of chromophore and one of radical), thus three dipolar interactions are possible: 1-2, 1-3 and 2-3 (where 1,2,3 represent the three electrons). However, due to the short distance between chromophore's electrons compared to that between the chromophore and the radical, only the coupling in between the two electron spins of the chromophore is taken into account.[1] This ZFS can then be expressed in the lab frame as

$$\hat{H}_{ZFS} = D_0[3\hat{S}_{z,c}^2 - S(S+1)] + D_{+1}(\hat{S}_{+,c}\hat{S}_{z,c} + \hat{S}_{z,c}\hat{S}_{+,c})$$
$$+ D_{-1}(\hat{S}_{-,c}\hat{S}_{z,c} + \hat{S}_{z,c}\hat{S}_{-,c}) + D_{+2}\hat{S}_{+,c}^2 + D_{-2}\hat{S}_{-,c}^2 \qquad \text{Eq. (S5)}$$

where

$$D_0 = \frac{D}{6}(3\cos^2\theta - 1) + \frac{E}{2}\sin^2\theta \cos 2\varphi$$

$$D_{\pm 1} = \frac{1}{4}\sin 2\theta(-D + E\cos 2\varphi) \pm i\frac{E}{2}\sin\theta \sin 2\varphi$$

$$D_{\pm 2} = \frac{1}{4}[D\sin^2\theta + E\cos 2\varphi(1 + \cos^2\theta)] \pm \frac{iE}{2}\cos\theta \sin 2\varphi$$

Here, $D$ and $E$ are the ZFS parameters of the chromophore and ($\theta$, $\phi$) represent the orientation of the ZFS tensor frame with respect to the Zeeman magnetic field. Since for anthraquinone molecule $D \gg E$,[2] we have assumed an axial symmetry for anthraquinone i.e., $E = 0$ for both the ANCOOT and Anq1Pr molecules.

For typical chromophore-TEMPO based systems (where TEMPO is linked to chromophore by a short linker), at X-band the magnitudes of the Hamiltonian terms can be ranked as followed:

$\|\hat{H}_{JCR}\|$ (1-5 cm$^{-1}$) > $\|\hat{H}_Z\|$ (0.3 cm$^{-1}$) ~ $\|\hat{H}_{ZFS}\|$ (0.3 cm$^{-1}$) >> $\|\hat{H}_{HF,14N}\|$ (1.4×10$^{-3}$ cm$^{-1}$)

While at high field (18.8 T, 527 GHz), the ranking is:

$\|\hat{H}_Z\|$ (~16 cm$^{-1}$) > $\|\hat{H}_{JCR}\|$ (1-5 cm$^{-1}$) >> $\|\hat{H}_{ZFS}\|$ (0.3 cm$^{-1}$) >> $\|\hat{H}_{HF,14N}\|$ (1.4×10$^{-3}$ cm$^{-1}$)

To simplify the RQM treatment, the hyperfine term which is very small compared to the other terms, is neglected.[1]

For the CR$_a$-R$_b$ in the fundamental state we have a time dependent Hamiltonian in the rotating frame defined as:

$$\hat{H}(t) = \hat{H}_Z(t) + \hat{H}_{HF,1H}(t) + \hat{H}_D(t) + \hat{H}_{JRaRb} + \hat{H}_{\mu w} \quad \text{Eq. (S6)}$$

where

$$\hat{H}_Z(t) = \sum_{i=a,b}(g_i(t)\beta_e B_0 - \omega_{\mu w})\hat{S}_{z,i} - \omega_n \hat{I}_{z,n}$$
$$\hat{H}_{HF,1H}(t) = A_{z,a,n}(t)\hat{S}_{z,a}\hat{I}_{z,n} + 2(A^+_{a,n}(t)\hat{S}_{z,a}\hat{I}^+_n + A^-_{a,n}(t)\hat{S}_{z,a}\hat{I}^-_n)$$
$$\hat{H}_D(t) = D_{a,b}(t)\left(2\hat{S}_{z,a}\hat{S}_{z,b} - \frac{1}{2}(\hat{S}^+_a\hat{S}^-_b + \hat{S}^-_a\hat{S}^+_b)\right) \quad \text{Eq. (S7)}$$
$$\hat{H}_{Jab} = -2J_{a,b}\left(\hat{S}_{z,a}\hat{S}_{z,b} + \frac{1}{2}(\hat{S}^+_a\hat{S}^-_b + \hat{S}^-_a\hat{S}^+_b)\right)$$
$$\hat{H}_{\mu w} = \sum_{i=a,b}\omega_1 \hat{S}_{x,i}$$

Herein, the electrons spins are designated by *a*, *b* while nuclear spins designated by *n*, $g_i$ stands for the electron g-tensor value of electron spin $i$, $D_{a,b}$, the electron dipolar coupling between electron spin *a* and *b*, $J_{a,b}$, the exchange interaction between electron spin *a* and *b*, $A_{a,n}$, the hyperfine coupling between electron spin *a* and proton nucleus *n*, $\omega_n$, the nuclear Larmor frequency of nucleus *n*, $\omega_{\mu w}$, the microwave frequency, and $\omega_1$, the microwave nutation frequency.

The electron polarization is defined as:

$$P_e = -2\text{tr}(\hat{\rho}\hat{S}_R). \quad \text{Eq. (S8)}$$

When the system is in thermal equilibrium (or Boltzmann equilibrium) at 100 K, we then obtain:

$$P_e^{eq} = -2\text{tr}(\hat{\rho}_{eq}\hat{S}_R) = 0.0023 \text{ at X band}$$
$$= 0.12 \text{ at } 18.8 \text{ T}/ 527 \text{ GHz}$$

The polarization of the nuclear spin *n* is written as $P_n$:

$$P_n = -2\mathrm{tr}(\hat{\rho}\hat{I}_z)$$

and at thermal equilibrium, $P_n^{eq}$, we obtain:

$$P_n^{eq} = -2\mathrm{tr}(\hat{\rho}_{eq}\hat{I}_z) = 2 \times 10^4 \text{ at } 18.8 \text{ T}/ 527 \text{ GHz}$$

Finally, the polarization gain $\epsilon_B$ is defined as:

$$\epsilon_B = \frac{P_n}{P_{n,eq}} \quad \text{Eq. (S9)}$$

We name "MAS-DNP field profile" the enhancement as a function of the magnetic field: $\epsilon_B = f(B_0)$. Note: we consider the microwave frequency fixed in this case.

**Appendix 2: Theoretical model for Fitting of the time resolved EPR (TREPR) time profile of ANCOOT at 100 K**

An understanding of the electron hyperpolarization (EHP) dynamics in the frozen solutions of ANCOOT (and Anq1Pr) at 100 K based on the reverse quartet mechanism (RQM) and the spin orbit induced intersystem crossing (SO-ISC) pathway would require two modifications from the existing approach (that is used for the solution state). i) The rate constant of $D_1$-$Q_1$ mixing, $k_{DQ}^{mn}$, of RQM pathway would become anisotropic as the perturbation responsible for $D_1$-$Q_1$ mixing, $H_{ZFS}$ (Eq. S5), is primarily anisotropic. In the solution state, an angular average is performed over the $k_{dq}$ values, which essentially removes the anisotropy (Eq. S11).[1] However, for the frozen conditions, this may not be strictly applicable. Moreover, knowledge of the relative orientation of the chromophore and the radical must be known. Only then it is possible to correlate the angles (θ, ϕ) of the zero-field splitting (ZFS) tensor of the chromosphere's triplet state (that governs the matrix elements of $H_{ZFS}$) with the angles corresponding to the $g/A$ tensor of the radical (that governs the actual powder EPR spectrum being observed). ii) The temperature dependence of $k_{DQ}^{mn}$ should be known to obtain its estimated value at 100 K.

A rigorous analysis addressing both of these points would digress us too far from the scope of the present article. For the present work, our aim is to obtain a simple estimate of the $k_{DQ}^{mn}$ at high fields and low temperature so that the efficiency of RQM could be gauged under these conditions. We thus followed a simplified approach to obtain an average picture applicable to glassy matrices (where all possible crystal orientations of a radical are found). The angular average of the $k_{DQ}^{mn}$ values was done in a similar way as that used in the solution state. This simplifies the first point which is associated with the anisotropic nature of $k_{DQ}^{mn}$. The second point was addressed by assuming that $k_{dq}$ follows an Arrhenius behavior, which is typically the case for other non-radiative processes like triplet ($T_1$) to singlet ($S_0$) state decay of photoexcited chromophores .[3] Thus, $k_{dq}$ at any given temperature is expressed using:

$$k_{DQ}^{mn}(T) = k_{DQ,RT}^{mn} e^{-E_a/RT} \qquad \text{Eq. (S10)}$$

where $E_a$ is the activation energy associated with the $D_1$-$Q_1$ process and $k_{DQ}^{mn}$ is the value of $k_{DQ,RT}^{mn}$ at room temperature and is given by Eq. S10. For simplicity, we assumed an identical $E_a$ for all the $D_1$-$Q_1$ transitions and obtained its value by numerical fitting of the X band TREPR time profile of ANCOOT at 100 K. The procedure is summarized below:

The $k_{DQ}^{mn}$ for the transition between the *m* and *n* sublevels of the $D_1$ and $Q_1$ state is given by[1]

$$k_{DQ}^{mn} = k_{DQ}^0 \times \frac{\overline{\langle D_1^m | \hat{H}_{ZFS} | Q_1^n \rangle^2}}{\Delta E(D_1^m, Q_1^n)^2} \qquad \text{Eq. (S11)}$$

where $k_{DQ}^0$ (ca. $10^{13}$ s$^{-1}$) is the product of the rate constant of zero-point motion and the Franck-Condon factor, $\Delta E$ is the energy difference between the *m* and *n* sublevels of the $D_1$ and $Q_1$ state, respectively, which depends on the Zeeman term, the exchange parameter $J_{CR}$, and the dipolar term, and $\hat{H}_{ZFS}$ is the ZFS perturbation. The bar over the $\hat{H}_{ZFS}$ matrix element in Eq. S11 represents an average over all angles between the magnetic field and the ZFS molecular frame.

A similar averaging was performed for the energy levels and the resulting $\Delta E_{DQ}$ values for various RQM transitions are shown in Table S1.

**Table S1.** $\Delta E_{DQ}$ associated with the RQM transitions between the $Q_1^m$, and $D_1^n$ substates.

| Transition | $\Delta E_{DQ}$ | Transition | $\Delta E_{DQ}$ |
|---|---|---|---|
| $Q_1^{-3/2} \rightarrow D_1^{+1/2}$ | $3|J_{CR}| - 2E_B$ | $Q_1^{-3/2} \rightarrow D_1^{-1/2}$ | $3|J_{CR}| - E_B$ |
| $Q_1^{-1/2} \rightarrow D_1^{+1/2}$ | $3|J_{CR}| - E_B$ | $Q_1^{+1/2} \rightarrow D_1^{-1/2}$ | $3|J_{CR}| + E_B$ |
| $Q_1^{+3/2} \rightarrow D_1^{+1/2}$ | $3|J_{CR}| + E_B$ | $Q_1^{+3/2} \rightarrow D_1^{-1/2}$ | $3|J_{CR}| + 2E_B$ |

Note: $E_B$ is the Zeeman term.

The rate equations governing the kinetic RQM model are defined as:

$$\frac{d[Q^{+3/2}]}{dt} = -\left\{ k_{qd}^{+3/2,-1/2} + k_{qd}^{+3/2,+1/2} + W_Q^{+3/2,-1/2} + W_Q^{+3/2,+1/2} \right.$$
$$\left. + k_{QD0} \right\}[Q^{+3/2}] + k_{dq}^{+1/2,+3/2}[D_1^{+1/2}]$$
$$+ k_{dq}^{-1/2,+3/2}[D_1^{-1/2}] + W_{-Q}^{+1/2,+3/2}[Q^{+1/2}] + W_{-Q}^{-1/2,+3/2}[Q^{-1/2}]$$

$$\frac{d[Q^{+1/2}]}{dt} = -\left\{ k_{qd}^{+1/2,-1/2} + W_{-Q}^{+1/2,+3/2} + W_Q^{+1/2,-1/2} + W_Q^{+1/2,-3/2} \right.$$
$$\left. + k_{QD0} \right\}[Q^{+1/2}] + k_{dq}^{-1/2,+1/2}[D_1^{-1/2}] + W_Q^{+3/2,+1/2}[Q^{+3/2}]$$
$$+ W_{-Q}^{-1/2,+1/2}[Q^{-1/2}] + k_{-Q}^{-3/2,+1/2}[Q^{-3/2}]$$

$$\frac{d[Q^{-1/2}]}{dt} = -\left\{ k_{qd}^{-1/2,+1/2} + W_{-Q}^{-1/2,+3/2} + W_{-Q}^{-1/2,+1/2} + W_Q^{-1/2,-3/2} \right.$$
$$\left. + k_{QD0} \right\}[Q^{-1/2}] + k_{dq}^{+1/2,-1/2}[D_1^{+1/2}] + W_Q^{+3/2,-1/2}[Q^{+3/2}]$$
$$+ W_Q^{+1/2,-1/2}[Q^{+1/2}] + k_{-Q}^{-3/2,-1/2}[Q^{-3/2}]$$

$$\frac{d[Q^{-3/2}]}{dt} = -\left\{ k_{qd}^{-3/2,-1/2} + k_{qd}^{-3/2,+1/2} + k_{-Q}^{-3/2,+1/2} + k_{-Q}^{-3/2,-1/2} + k_{QD0} \right\}[Q^{-3/2}]$$
$$+ k_{dq}^{+1/2,-3/2}[D_1^{+1/2}] + k_{dq}^{-1/2,-3/2}[D_1^{-1/2}] + W_Q^{+1/2,-3/2}[Q^{+1/2}]$$
$$+ W_{-Q}^{-1/2,-3/2}[Q^{-1/2}]$$

Eq. (S12)

$$\frac{d[D_1^{+1/2}]}{dt} = -\left\{ k_{dq}^{+1/2,+3/2} + k_{dq}^{+1/2,-1/2} + k_{dq}^{+1/2,-1/2} + W_D + k_{qt}^{+1/2,+1/2} \right\} [D_1^{+1/2}]$$
$$+ k_{qd}^{+3/2,+1/2}[Q^{+3/2}] + k_{qd}^{-1/2,+1/2}[Q^{-1/2}] + k_{qd}^{-3/2,+1/2}[Q^{-3/2}]$$
$$+ W_{-D}[D_1^{-1/2}]$$

$$\frac{d[D_1^{-1/2}]}{dt} = -\left\{ k_{dq}^{-1/2,+3/2} + k_{dq}^{-1/2,+1/2} + k_{dq}^{-1/2,-3/2} + W_{-D} + k_{qt}^{-1/2,-1/2} \right\} [D_1^{-1/2}]$$
$$+ k_{qd}^{+3/2,-1/2}[Q^{+3/2}] + k_{qd}^{+1/2,-1/2}[Q^{+1/2}] + k_{qd}^{-3/2,-1/2}[Q^{-3/2}]$$
$$+ W_D[D_1^{+1/2}]$$

$$\frac{d[D_0^{+1/2}]}{dt} = k_{qt}^{+1/2,+1/2}[D_1^{+1/2}] - W_D[D_0^{+1/2}] + W_{-D}[D_0^{-1/2}]$$

$$\frac{d[D_0^{-1/2}]}{dt} = k_{qt}^{-1/2,-1/2}[D_1^{-1/2}] + W[D_0^{+1/2}] + W_{-D}[D_0^{-1/2}]$$

The solution of these differential equation is used to obtain the population difference between the $D_0^{+1/2}$ and $D_0^{-1/2}$ sublevels, which is a measure of hyperpolarization in the ground state of CR molecule. The time dependent EHP curve was convolved with the instrument response time.

To this end, a knowledge of various parameters such as the initial population distribution in $D_1$, $Q_1$ states (just after the laser excitation), the spin-lattice relaxation (SLR) rates, $W_{D1}$, $W_{Q1}$, and $W_{D0}$ in the $D_1$, $D_0$, and $Q_1$ states, respectively, the rate constant for SO-ISC process ($k_{ISC-SO}$), the quenching process ($k_{qt}$), and the $D_1$-$Q_1$ transfer via RQM ($k_{DQ}^{mn}$) and the $E_a$ value of $k_{DQ}^{mn}$ are needed.

To simplify the kinetics (and to minimize the number of variable parameters), the following assumptions were made:

i. The initial EHP in the $D_1$ and $Q_1$ state was assumed to be governed by the SO-ISC process while its subsequent evolution was governed by RQM. This assumption was based on the large difference in the corresponding rate constants at RT; via. $k_{SO-ISC} \sim 10^{12}$ s$^{-1}$,[4] and $k_{dq} \sim 10^6$-$10^9$ s$^{-1}$ (depending on the $D$ value of chromophore and $\Delta E$).[1,5] Though these rate constants are expected to decrease at 100 K, the ISC process for molecules with such an ultrafast nature are expected to show a very weak temperature dependence. This is evident from the fact that the structurally similar molecule benzophenone, which is also known to possess an ultrafast ISC at RT, shows almost complete absence of fluorescence even at a low temperature of 77 K (indicating the dominance of ISC at LT too).[2] This behavior is rationalized to the very small activation energy for the ISC process in such molecules (as a result of the almost degeneracy between the $S_1$ and $T_2$ states between which the ISC occurs).[3,6,7] Furthermore, we assumed the initial population (generated via SO-ISC) only in the $Q_1^{+3/2}$ and $D_1^{+1/2}$ state i.e., $Q_1^{+3/2}=1$, $Q_1^{+1/2}=0$, $Q_1^{-1/2}=0$, $Q_1^{-3/2}=0$, $D_1^{+1/2}=b$, $D_1^{-1/2}=0$, where the value of $b$ was obtained from best fit. This assumption was based on the anthraquinone's behavior to generate emissively polarized radicals in its photochemical reactions via the triplet state,[8] thereby indicating a preferential population of $T_{+1}$ triplet state during its ISC (Note, $Q_1^{+3/2}$ and $D_1^{+1/2}$ represent the states that have the largest contribution of $T_{+1}$ state in the $Q_1$ and $D_1$ manifold, respectively).[9]

ii. The SLR rates in $D_1$ and $D_0$ state were assumed to be same. This assumption is supported by their earlier estimated SLR rates in the fullerene-TEMPO system.[1]

iii. A $J$ value of $-3.4$ cm$^{-1}$ was assumed for ANCOOT, which was based on the earlier reported $J$ value for a fullerene-TEMPO complex. Our choice of fullerene-TEMPO molecule as a reference for $J$ was based on the identical nature of the radical (i.e., TEMPO), as well as to similarity of the linker (ester linkage) and the C-R separation in the two molecules.[1]

iv. The rate constant for the intrinsic decay of anthraquinone from the $Q_1$ to $D_0$ state, $k_Q^0$, was assumed to have a value of 303 s$^{-1}$. This was based on the reported anthraquinone's phosphorescence lifetime of 3.3 ms at 77 K.[2]

The values of the remaining parameters were obtained from the simulations. The range over which their values were varied was obtained as follows:

$W_{Q1}$: The dominant relaxation mechanism for the organic triplet state molecules in solid state has been found to be the modulation of electron spin dipolar Hamiltonian by the lattice phonons (termed the direct process).[10,11] The SLR rate of a direct process between two spin states separated in energy by δ is given by

$$W = \frac{3\delta^2 kT |\langle a|H'|b\rangle|^2}{\pi h^4 v_s^5 \rho}$$

Eq. (S13)

where $H'$ is the electron-electron dipolar Hamiltonian ($H_{dd}$), $v_s$ is the velocity of sound in the medium and $\rho$ is the density of the medium. Thus, the above equation predicts a quadratic dependence of $W_{Q1}$ on the $D$ value of chromophore.

Next, we used the reported SLR rate of pentacene's triplet state to obtain a corresponding estimate of anthraquinone's SLR rate as follows: The SLR rate of pentacene averaged over all three principle directions and the two triplet transitions in p-terphenyl host at 100 K is ~ 3.3 ± 2.3×10$^4$ s$^{-1}$.[11] Using the reported $D$ values of anthraquinone, 0.29 cm$^{-1}$ in octane (and 0.35 cm$^{-1}$ in diethylether/isopentane/ethanol (5:5:2 by volume) = EPA), and pentacene, 0.046 cm$^{-1}$ in p-terphenyl, and following Eq. S4, a $W_{Q1}$ value of ~ 1×10$^6$ s$^{-1}$ (2×10$^6$ s$^{-1}$) is estimated for anthraquinone's triplet state at 100 K in octane (and EPA). Similarly, a value of 0.1×10$^6$ s$^{-1}$ has been measured for the triplet SLR time of benzophenone at 30 K.[12] Thus, at 100 K a value of 0.9×10$^6$ s$^{-1}$ is estimated for anthraquinone.

Based on these values, a $W_{Q1}$ range of 0.05-3×10$^6$ s$^{-1}$ was used for the anthraquinone molecule at 100 K.

$k_{qt}$: A $k_{qt}$ value of 10$^8$ s$^{-1}$ was used in the earlier solution-state RQM simulations of ANCOOT molecule.[13] This value was based on the measured $k_{qt}$ value for the structurally similar thioxanthone-TEMPO CR molecule.[14] In the present study, the rise time of the emissive EHP in ANCOOT and Anq1Pr recorded at 100 K was almost identical with the one recorded at 300 K (solution state experiment). We thus used this value as the upper limit of $k_{qt}$. The lower bound is equal to the rise time of the instrument response time ($k_{IRF}$ ~ 10$^7$ s$^{-1}$) of the TREPR spectrometer. Thus, $k_{qt}$ was varied between 10$^7$ and 10$^8$ s$^{-1}$ until a best fit was obtained.

$k_{DQ}^{mn}$: it can be estimated based on Eq. S10. To obtain a best fit $E_a$ value from the simulations, $E_a$ was increased from (0 KJ/mol) until a good fit was obtained.

$W_{D0}$: the $W_{D0}$ of ANCOOT molecule was obtained by pulsed saturation recovery at the same temperature where the electron spin polarization (ESP) was recorded (100 K). This gave a value of 78±5 and 65±5 μs at position $x_1$ and $x_2$, respectively for the ANCOOT molecule in toluene at 100 K (see Fig. S4 for the definition of $x_1$ and $x_2$). In the simulations, these values were used as the initial guesses.

Under these conditions, the coupled differential equations (Eq. S12) were numerically solved using a fourth order Runge-Kutta method, from which the population difference between the two spin sublevels of the ground $D_0$ state was obtained. To make comparisons easier, the difference in $D_0$ spin population, i.e., $D_0^{+1/2}$–$D_0^{-1/2}$ were normalized with respect to the equilibrium population difference of the same two levels. In the plots (Fig S1), an offset of −1 is added, such that the signal is referenced to zero.

Fig. S1 presents the comparison between the simulation and the experimental TREPR time profile of ANCOOT at 100 K, which showed that the simulations reasonably reproduced the observed electron hyperpolarization in the entire time window. The best fit $E_a$ value was found to be 10.3±0.3 kJ/mol which resulted in a $k_{dq}$ value of ~5×10$^3$ s$^{-1}$ for the $Q_1^{-3/2}$ to $D_1^{+1/2}$ transition. This indicated that the RQM pathway occurs on a much slower timescale compared to the SO-ISC pathway at X-band frequency and frozen conditions (100 K). Thus, the initial fast rise of the hyperpolarization is due to the rapid transfer of the hyperpolarization from the $D_1$ state (generated via the SO-ISC pathway) to the $D_0$ state via the quenching process ($k_{qt}$ ~ 20 × 10$^6$ s$^{-1}$), while the subsequent evolution is governed by the RQM pathway and the SLR process of the radical (in the $D_0$ state). The slow nature of RQM pathway was more evident when we performed the numerical simulations by incorporating only the SO-ISC contribution. This was achieved by zeroing the $k_{dq}$ value (which is done by using a $D$ value of zero). Note that in that case, only the polarized spin population in $D_1$ state would be transferred to the $D_0$ state via SO-ISC. In addition, since the $Q_1$-$D_1$ pathway is blocked, the $Q_1$ state molecules would return to the ground state via the non ESP generating $Q_1 \rightarrow D_0$ pathway ($k_Q^0$). Interestingly, a pure SOISC pathway could not fit the experimental curve for any reasonable set of simulation parameters. A good fit was seen only in the initial time window, where SO-ISC contribution was dominant.

The above simulations support the model with a fast electron hyperpolarization generation via the SO-ISC process followed by a slow generation via the RQM pathway in the ANCOOT (and Anq1PR molecules) at X-band and frozen state (low temperature).

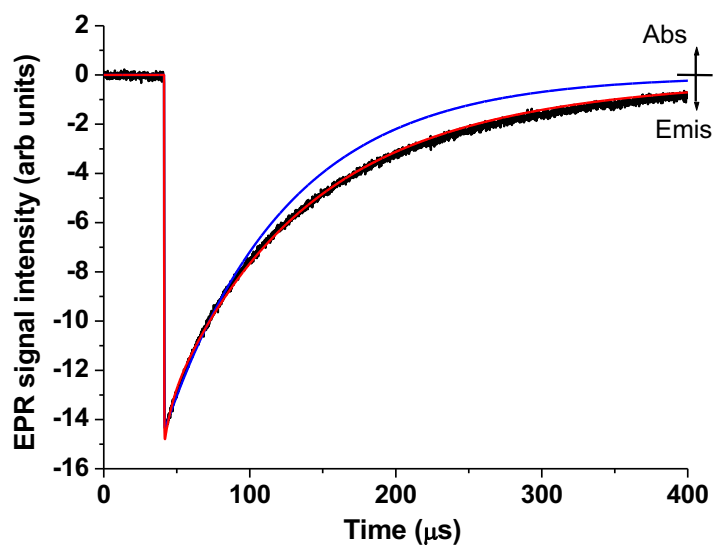

**Figure S1.** Comparison between the experimental (black) TREPR time profile of ANCOOT in toluene recorded at the position $x_1$ (see fig. S4) and its simulation that considers only SO-ISC pathway (blue) and SO-ISC + RQM pathway (red) : The best fit simulation parameters for the two cases were as follows: Only SO-ISC: $D = 0$; $W_{D0} = 0.0062\times10^6$ s$^{-1}$, $W_{Q1} = 0.1\times10^6$ s$^{-1}$, $k_{qt} = 20\times10^6$ s$^{-1}$, $k_Q^0 = 303$ s$^{-1}$, Initial population: $Q_1^{+3/2} = 1$, $Q_1^{+1/2} = 0$, $Q_1^{-3/2} = 0$, $Q_1^{-1/2} = 0$, $D_1^{+1/2} = 0.75$, $D_1^{-1/2} = 0$, and B) SO-ISC + RQM pathway: $D = 0.31$ cm$^{-1}$ ; $W_{D0} = W_{D1} = 0.0062\times10^6$ s$^{-1}$, $W_{Q1} = 0.1\times10^6$ s$^{-1}$, $k_{qt} = 20\times10^6$ s$^{-1}$, $E_a = 10.4$ KJ/mol, Initial population: $Q_1^{+3/2} = 1$, $Q_1^{+1/2} = 0$, $Q_1^{-3/2} = 0$, $Q_1^{-1/2} = 0$, $D_1^{+1/2} = 0.22$, $D_1^{-1/2} = 0$. Note: a gain of 2.5 was used for the simulated curve in the SO-ISC+RQM model. The mismatch in the long-time region in pure SO-ISC model is reduced substantially in SO-ISC+RQM. Temperature: 100 K. **"Abs"** and **"Emis"** denote the absorptive and emissive spin polarization.

**Appendix 3: *J* dependence of the electron hyperpolarization generated via the RQM pathway**

The magnitude of the electron hyperpolarization generated via the RQM pathway depends not only on the efficiency of the $D_1$-$Q_1$ mixing,[1,5] but also on its efficiency towards a *selective* population of the $D_1$ spin sublevels. The former would depend on the magnitude of $k_{dq}$ as governed by equation S2, while the latter would primarily be governed by the ratio of the total rates which populates the two $D_1$ spin sublevels. Importantly, the energy gap, $\Delta E_{DQ}$, between the $D_1$ and $Q_1$ sublevels governs both the magnitude of $k_{dq}$ as well as the preferential population of $D_1$ sublevels (as $\Delta E_{DQ}$ is different for each $D_1$-$Q_1$ transition). Furthermore, since $\Delta E_{DQ}$ has a dominant contribution from the exchange ($J_{CR}$) interaction,[1] $J_{CR}$ would likely play a strong role in governing the RQM efficiency.

First we calculate the sum of all the rate constants that connects the $Q_1$ spin substates to the $D_1^{+1/2}$ and to the $D_1^{-1/2}$ substates via the RQM pathway. To simplify the calculations (and remove the anisotropy associated with the dipolar Hamiltonian) we considered the isotropic average of the $k_{dq}$ value (Eq. S11). Its value at 100 K was estimated based on the activation energy obtained from the numerical simulation of the X band EPR TREPR time profile of ANCOOT molecule at 100 K (Appendix 2). The ratio of these sums gives a quantity, $R_{D1}$, which we define as the selectivity factor. Specifically, $R_{D1}$ can be written as:

$$R_{D1} = \frac{k_{dq}^{Q_1^{+3/2}D_1^{+1/2}} + k_{dq}^{Q_1^{-1/2}D_1^{+1/2}} + k_{dq}^{Q_1^{-3/2}D_1^{+1/2}}}{k_{dq}^{Q_1^{+3/2}D_1^{-1/2}} + k_{dq}^{Q_1^{+1/2}D_1^{-1/2}} + k_{dq}^{Q_1^{-3/2}D_1^{-1/2}}} \quad \text{Eq. (S14)}$$

We use rate constants, $k_{DQ}^{mn}$, instead of the rates $k_{DQ}^{mn} \times [Q_1^m]$ to evaluate the selectivity factor in order to mimic a pure RQM situation, where the initial $Q_1$ state population is governed by its thermal equilibrium ESP; RQM does not need the $Q_1$ state to be polarized initially. Furthermore, since the thermal equilibrium ESP is << 1 at 100 K, the population in the $Q_1$ spin sublevels will almost be equal (i.e., their relative populations can be taken to be $Q_1^{+3/2} = 1$, $Q_1^{+1/2} = 1$, $Q_1^{-3/2} = 1$, $Q_1^{-1/2} = 1$). Thus, $[Q_1^m]$ being the same for all $Q_1$ sublevels at initial time (i.e., just after the laser pulse), the corresponding population of $D_1$ states via the $Q_1$-$D_1$ RQM transitions would be determined by the rate constant. Hence the use of rate constant becomes justified and $R_{D1}$ can be taken as a measure of preferential population of $D_1$ spin substates and thereby of the RQM efficiency.

Using the obtained $R_{D1}$ value, we calculate the electron hyperpolarization in the $D_1$ substates (and consequently in the $D_0$ substates) as follows:

$$P_i^{pol} = \frac{D_1^{-\frac{1}{2}} - D_1^{+\frac{1}{2}}}{D_1^{-\frac{1}{2}} + D_1^{+\frac{1}{2}}} = \frac{D_1^{-\frac{1}{2}} - R_{D1}D_1^{-\frac{1}{2}}}{D_1^{-\frac{1}{2}} + R_{D1}D_1^{-\frac{1}{2}}} = \frac{1 - R_{D1}}{1 + R_{D1}} \quad \text{Eq. (S15)}$$

Based on Eq. S8, it is easy to see that a $R_{D1}$ value greater than 10 would result in a $P_i^{Pol}$ value (i.e., the ESP in the $D_1$ and $D_0$ states) greater than 80%. Table S2 and fig. 1B shows the values of the $R_{D1}$ and the corresponding $P_i^{Pol}$ values at both $B_0 = 0.33$ T/$\nu = 9.4$ GHz (X-band) and $B_0 = 18.8$ T/ $\nu = 527$ GHz (where $\nu$ is the EPR resonance frequency). It is evident that large $R_{D1}$ values are observed for a wide range of $J$ values ~ 12.0±2.5 cm$^{-1}$ thereby giving a broad $J$-window of ~5 cm$^{-1}$. This is quite reasonable to achieve for CR molecules (for example by adjusting the CR separation and or their relative orientation).

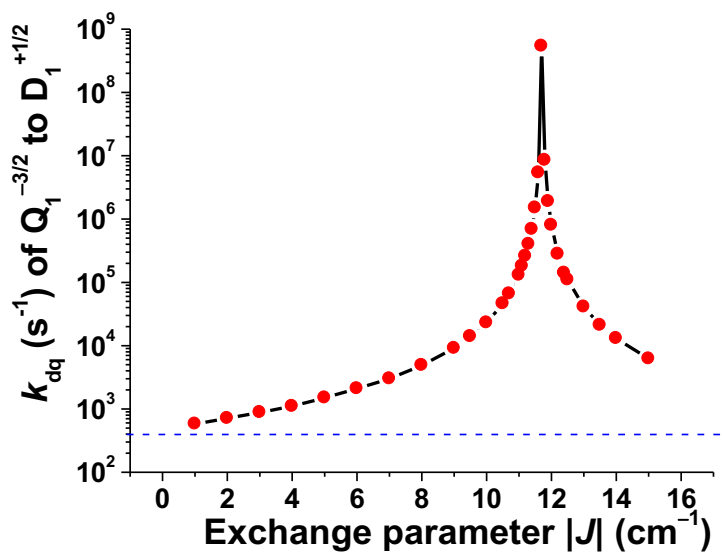

**Figure S2**: Plot of $k_{dq}$ value for $Q_1^{-3/2} \to D_1^{+1/2}$ transition at 18.8 T as a function of exchange parameter, $J$. The blue dashed line points to the intrinsic decay constant $k_Q^0$ of the $Q_1$ state. The black solid line is a guide to the eye.

Notably, the $k_{dq}$ value for $Q_1^{-3/2} \to D_1^{+1/2}$ transition also becomes significantly larger than the intrinsic decay constant of the $Q_1$ state, $k_Q^0$ (~$3 \times 10^2$ s$^{-1}$), in the broad $J$-window (fig. S2). This is essential to generate a large ESP, since the large $k_{dq}$ value would enable the $D_1$-$Q_1$ pathway to compete efficiently with the non ESP generating $Q_1 \to D_0$ pathway.

Importantly, this electron hyperpolarization can be generated without any contribution of the SO-ISC pathway i.e., only via the RQM pathway (since we assumed an equal $Q_1$ state populations at the initial time). As pointed in the main text, this result becomes more significant when one notes that the efficiency of SO-ISC pathway degrades at high fields. Specifically, the efficiency of SO-ISC depends on the generation of electron spin polarization (ESP) in the triplet state of the chromophore during its ISC. Importantly, the ISC occurs in the zero-field triplet levels ($T_x$, $T_y$, $T_z$) while the net polarization transferred to the radical originates from the net spin polarization in the Zeeman field triplet levels ($T_{+1}$, $T_0$, $T_{-1}$) of the chromophore. The relative populations ($P_0$, $P_{+1}$, $P_{-1}$) in the $T_{+1}$, $T_0$, $T_{-1}$ are given by [15,16]

$$P_0 = \frac{1}{3}(k_X + k_Y + k_Z)$$
$$P_{\pm 1} = P_0 \pm \left(2/5\right)\frac{D_T}{B_0}(k_X + k_Y + k_Z)$$

Eq. (S16)

where $k_x$, $k_y$, and $k_z$ are the corresponding relative ISC rates of the "zero-field" levels. From Eq. 1, it is evident that the net polarization in the $T_{+1}$, $T_0$, $T_{-1}$ levels has an inverse field dependence and hence it would be transferred to the radical will also become inefficient at high fields. In fact, it is expected that the SO-ISC pathway *could* decrease by a factor ~60 going from 0.33 to 18.8 T. An important conclusion of this work is that in the high field condition, only the RQM process is capable of generating a large electron spin hyperpolarization. As a passing remark, the efficiency of widely used triplet state DNP would also decrease at high fields if it is the net ESP of the triplet state (Eq. S16) that is dominantly responsible for the DNP enhancement in the triplet state DNP.

We wish to point here that the above treatment assumes the initial population in only the $Q_1$ state. In principle, the SO-ISC would govern the initial populations in the $Q_1$ and $D_1$ state (i.e., immediately after the laser irradiation). If the SO-ISC is inefficient at high Zeeman fields (see above), then the initial populations in the $Q_1$ and $D_1$ state would tend to become equally distributed (i.e., population of each $D_1$ and $Q_1$ sublevel would be 1/6). Since the equally populated $D_1$ state population would be lost to $D_0$ state without generating any ESP, the maximum ESP that one can obtain would be from the $Q_1$ state population via RQM. Thus, after a single laser pulse, the maximum ESP would be 4/6 (i.e., 67 %), not 100%. However, RQM is a slow process ($10^4$-$10^7$ s$^{-1}$) compared to the SO-ISC mediated $D_2$-$D_1$-$D_0$ pathway (both $k_{ISC}$ and $k_{qt}$ >$10^7$ s$^{-1}$). Thus, at a time interval of ~ $k_{qt}^{-1}$ after the laser pulse, the dominant population in the $D_0$ state would be the unpolarised (2/6 i.e., 33%) component. If another laser pulse is incident around this time window (or the pulse width of the first pulse itself is of the order of $k_{qt}^{-1}$), then this unpolarised population can be repumped into the $Q_1$/$D_1$ manifold. In other words, it should be possible to pump almost the entire population in the $Q_1$ state and hence extract 100% ESP via RQM. The entire cycle is summarized in scheme S1. This mechanism is what we name "optically pumped electron spin hyperpolarization".

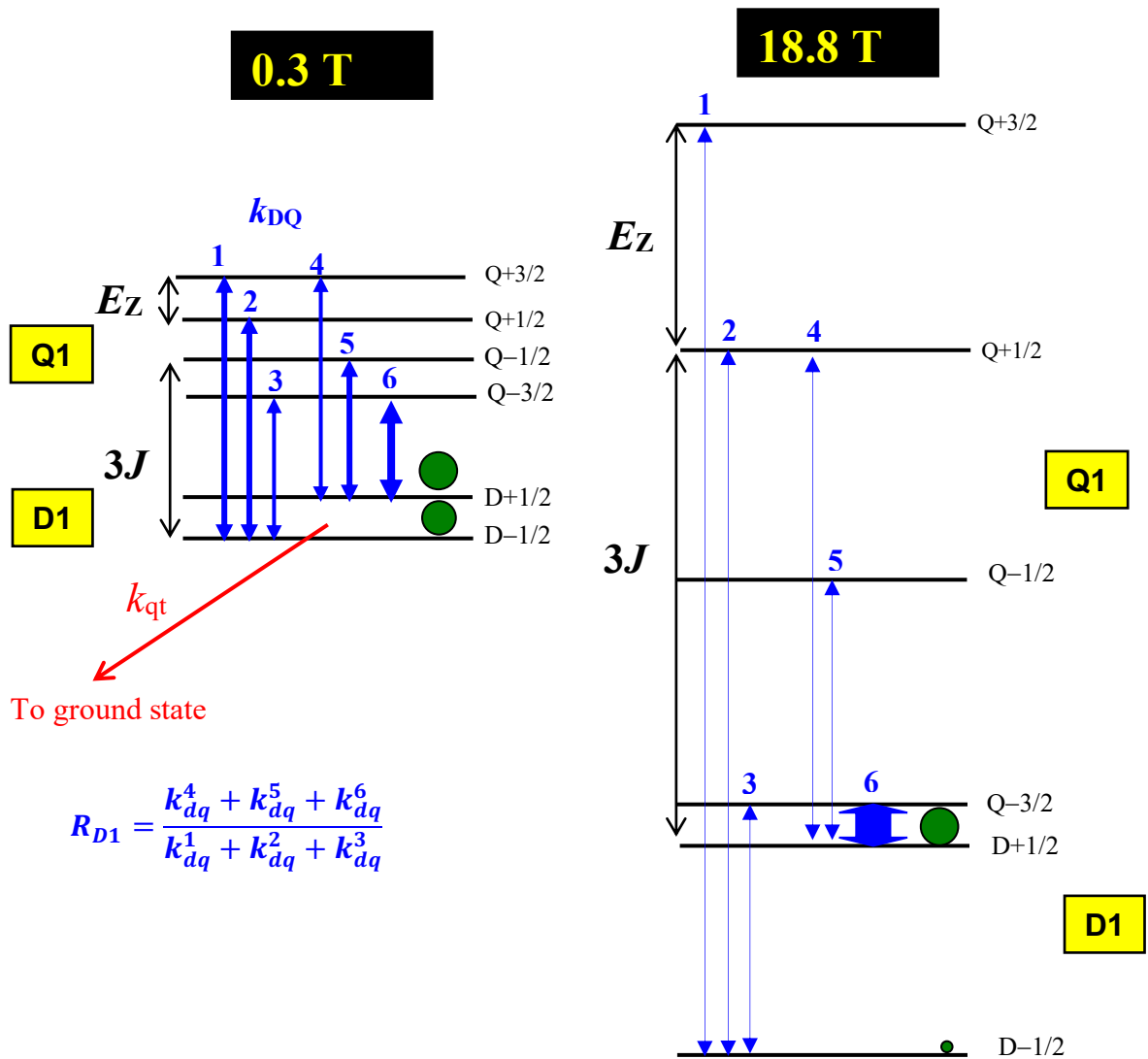

**Figure S3**: $D_1$-$Q_1$ energy levels of a C-R molecule depicting the RQM transitions between the $D_1$ and $Q_1$ levels (blue double headed arrows) at two different field strengths. The thickness of the arrow represents the magnitude of the $k_{dq}$ rate constant. $E_Z$ represents the Zeeman energy while $3J$ represents the magnitude of the exchange splitting. The green circles denote the population of $D_1$ states after RQM process. At high magnetic fields (18.8 T), i.e. large Zeeman interaction, the rate constant for the $D_1^{+1/2}$ and $Q_1^{-3/2}$ (transition no. 6) becomes the most dominating factor in governing the net $D_1$-$Q_1$ transition process. This results in a large overpopulation in the $D_1^{+1/2}$ state compared to $D_1^{-1/2}$ state resulting in an emissive ESP in $D_1$ and hence $D_0$ state.

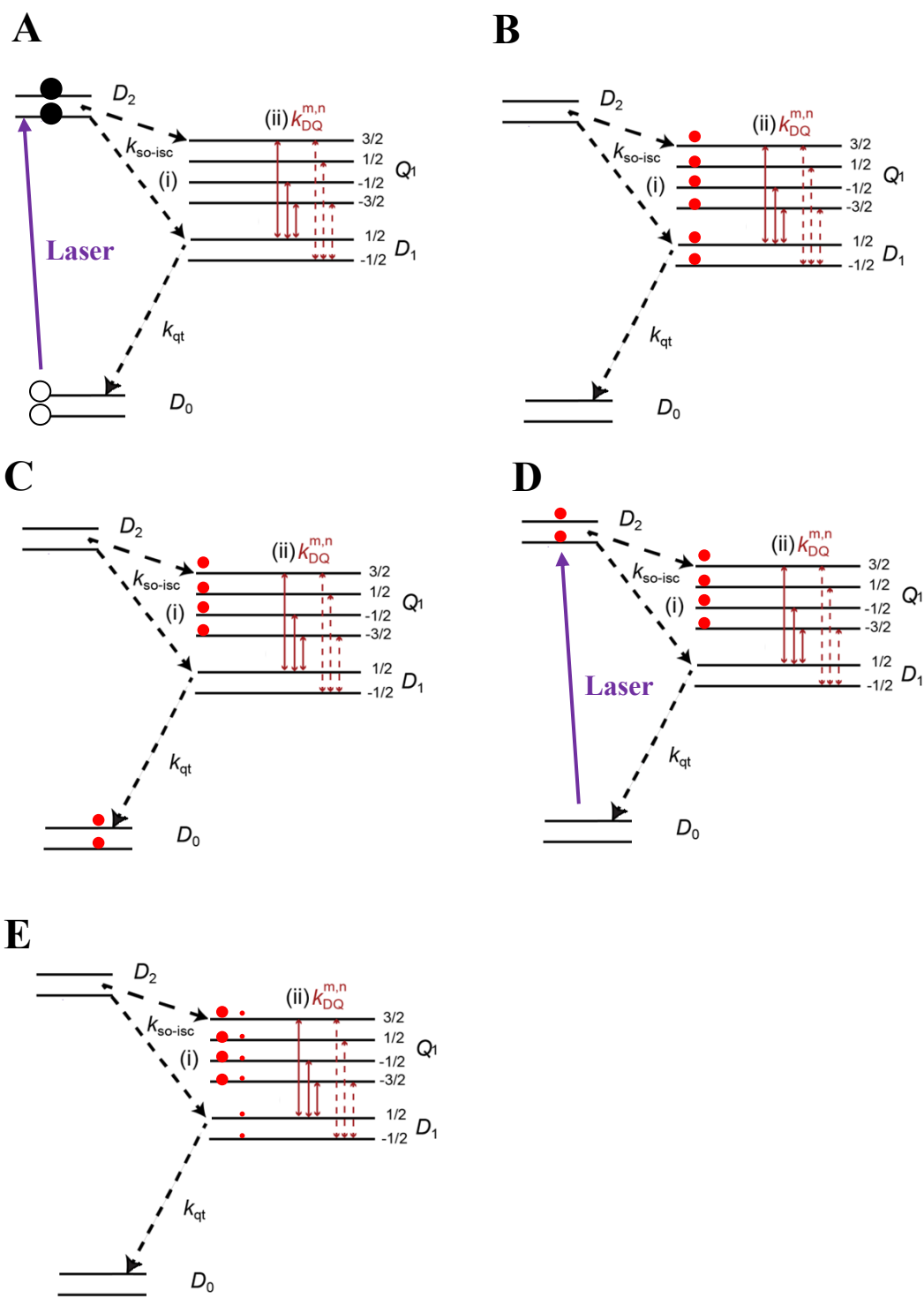

**Scheme S1**: Schematic of the optical pumping of $D_0$ state population into $Q_1$ state. Energy level diagram of the radical-chromophore system A) after the initial photoexcitation by a laser pulse. The population before the laser pulse is shown by hollow spheres. B) After the SO-ISC process, C) Quenching of the $D_1$ state to the ground $D_0$ state, D) Retransfer of the $D_1$ state population by the second laser pulse, E) After the SO-ISC process. Note, the SO-ISC process is assumed to be completely inefficient in generating an ESP in the $D_1Q_1$ manifolds.

**Table S2**: $R_{D1}$ values, $k_{dq}$ value for $Q_1^{-3/2} \rightarrow D_1^{+1/2}$ transition (denoted by $k_{DQ}{}^a$) and the initial ESP calculated for two different EPR frequency, $\nu = 9.5$ GHz ($B_0 = 0.33$ T) and 527 GHz ($B_0 = 18.8$ T) as a function of $J$. $E_a = 10.2$ kJ/mol, $k_{qt} = 20 \times 10^6$ s$^{-1}$, $D = 0.31$ cm$^{-1}$. $R_{D1}$ values which result in an initial ESP of greater than 80% are highlighted in red.

| B=0.38T/$\nu$ = 9.5 GHz | | | | B = 18.8T/$\nu$ = 527 GHz | | | |
|---|---|---|---|---|---|---|---|
| $J$(cm$^{-1}$) | $R_{D1}$ | $k_{DQ}{}^a$(s$^{-1}$) | $P_i{}^{pol}$ | $J$(cm$^{-1}$) | $R_{D1}$ | $k_{DQ}{}^a$(s$^{-1}$) | $P_i{}^{pol}$ |
| −0.1 | 3 | 5.5×10⁶ | -0.5 | -1 | 1.4 | 5.8×10² | -0.17 |
| −0.2 | 236 | 6.57×10⁸ | -0.99 | -2 | 1.9 | 7.1×10² | -0.31 |
| −0.3 | 9.67 | 8.3×10⁶ | -0.81 | -3 | 2.4 | 8.8×10² | -0.41 |
| −0.4 | 4.39 | 1.9×10⁶ | -0.63 | -4 | 2.8 | 1.1×10³ | -0.47 |
| −0.5 | 3.1 | 8.0×10⁵ | -0.51 | -5 | 3 | 1.5×10³ | -0.5 |
| −0.6 | 2.5 | 4.4×10⁵ | -0.43 | -6 | 3 | 2.1×10³ | -0.5 |
| −0.7 | 2.17 | 2.8×10⁵ | -0.37 | -7 | 3.1 | 3.0×10³ | -0.51 |
| −0.8 | 1.96 | 1.9×10⁵ | -0.32 | -8 | 3.9 | 4.9×10³ | -0.59 |
| −0.9 | 1.81 | 1.4×10⁵ | -0.29 | -9 | 6.9 | 9.1×10³ | -0.75 |
| −1 | 1.71 | 1.1×10⁵ | -0.26 | -9.5 | 10.8 | 1.4×10⁴ | -0.83 |
| −2 | 1.3 | 2.1×10⁴ | -0.13 | -10 | 19.6 | 2.3×10⁴ | -0.9 |
| −3 | 1.19 | 8.6×10³ | -0.09 | -10.5 | 43.6 | 4.6×10⁴ | -0.96 |
| −4 | 1.14 | 4.7×10³ | -0.07 | -10.7 | 65.7 | 6.6×10⁴ | -0.97 |
| −5 | 1.11 | 2.9×10³ | -0.05 | -11 | 143.1 | 1.3×10⁵ | -0.99 |
| −5.5 | 1.1 | 2.4×10³ | -0.05 | -11.1 | 198.6 | 1.8×10⁵ | -0.99 |
| −6 | 1.09 | 2.0×10³ | -0.04 | -11.2 | 291.3 | 2.6×10⁵ | -0.99 |
| | - | | - | -11.3 | 461.7 | 4.0×10⁵ | -1 |
| | - | | - | -11.4 | 900 | 6.9×10⁵ | -1 |
| | - | | - | 11.5 | 1842.2 | 1.5×10⁶ | -1 |
| | - | | - | -11.6 | 6821.4 | 5.4×10⁶ | -1 |

|  | - |  | - | -11.7 | 700000 | $5.4\times10^8$ | -1 |
|--|---|--|---|-------|--------|-----------------|----|
|  |   |  |   | -11.8 | 11000  | $8.5\times10^6$ | -1 |
|  |   |  |   | -11.9 | 2546.3 | $1.9\times10^6$ | -1 |
|  |   |  |   | -12   | 1116.2 | $8.0\times10^5$ | -1 |
|  |   |  |   | -12.2 | 409.6  | $2.8\times10^5$ | -1 |
|  |   |  |   | -12.4 | 217    | $1.4\times10^5$ | -0.99 |
|  |   |  |   | -12.5 | 169.3  | $1.1\times10^5$ | -0.99 |
|  |   |  |   | -13   | 71.4   | $4.1\times10^4$ | -0.97 |
|  |   |  |   | -13.5 | 41.5   | $2.1\times10^4$ | -0.95 |
|  |   |  |   | -14   | 28.2   | $1.3\times10^4$ | -0.93 |
|  |   |  |   | -15   | 16.7   | $6.2\times10^3$ | -0.89 |

**Appendix 4: Qualitative method to obtain a lower estimate of the $P_{POL}/P_{eq}$ for ANCOOT from the maximum amplitude of the time resolved EPR (TREPR) signal**

To determine the magnitude of electron spin polarization (ESP) enhancement (i.e., $P_{Pol}/P_{eq}$ where $P_{Pol}$ and $P_{eq}$ are the ESP of polarized and thermal equilibrium signal respectively) at low temperature and frozen conditions, some challenges had to be overcome. One of the major challenge was that the conventional ESP determination protocol (used in the solution state), which is based on the numerical fitting of the experimental TREPR profile to the kinetic Bloch equations,[5,17,18] could not be adapted to the frozen sample. This is primarily because the solution state approach is only applicable to homogenously broadened line, while at frozen conditions, the strong inhomogeneous broadening of the nitroxide EPR spectrum makes the application of same routine cumbersome. Thus, we resorted to a qualitative method to obtain an estimate of ESP. This approach was used in prior work for solution state experiments of both ANCOOT and Anq1Pr to obtain a lower bond of ESP magnitude.[5,13] The salient features of the qualitative method are as follows:

i) The TREPR transient profile of the sample is collected such that the EPR signal before and after the laser pulse gives a measure of the thermal equilibrium signal and the ESP signal, respectively (fig. S5A). The polarized signal is characterized by a rising part (governed by ESP generation) and a decaying part (governed by spin relaxation processes). If the relaxation processes had been negligible, then the observed polarized signal amplitude, should directly give the ESP enhancement with respect to the thermal equilibrium signal, provided all the molecules in the active volume (i.e., the laser irradiated sample volume) have absorbed light. Note: the typical laser energy used in the experiments leads to an excited molecule concentration of, $S_1^0$ ~ 10% of the ground state concentration, $R_{GND}$. However, due to finite relaxation processes, this method gives only a lower bound of the estimated ESP value.

ii) It is important that $P_{eq}$ should correspond to the thermal equilibrium polarisation of only those molecules which can generate ESP, i.e., which are exposed to the laser irradiation, and not the entire sample present in the microwave cavity. Only then can the ratio $P_{POL}/P_{eq}$ be used as a quantitative measure of the change of polarisation brought by photoexcitation of the chromophore. In an EPR experiment, the observed thermal equilibrium signal corresponds to all the radicals present in the cavity, however, ESP is generated only in the region where the laser pulse excites the sample (Fig. S5A). The remaining radicals do not participate in the ESP generation process, and hence it is necessary to subtract their contribution from the thermal equilibrium signal. This correction was performed first and is shown in Fig. S5B.

iii) The EPR signal is then normalized with respect to the Boltzmann signal (fig. S5C), and then normalized again with respect to the excited state concentration, $S_1^0$. This gives us a lower bound of the $P_{Pol}/P_{eq}$ factor.

It is important to point here that the determination of ESP enhancement (either via Bloch equation approach or the qualitative method) requires an accurate reproduction of both the polarized and the Boltzmann signal (under identical conditions). In typical TREPR spectrometers, the Boltzmann signal (being time independent) is filtered out by the preamplifier. To overcome this rejection, we vary the magnetic fields (i.e. field jumps) alternatively between on and off-resonance in sync with the pulsed laser frequency.[5,19] For each on- and off-resonances, the TREPR time profile is recorded and then the difference of the two transients is used. Importantly, in this routine, it is crucial to select an off-resonance position which is free from any resonance signal. In the solution state experiments, this is an easy task since even a 5 G off resonance jump from the resonance position (line width~1 G) results in a clean off-resonance region. However, in case of a frozen-state sample, the presence of inhomogeneous broadening causes the spectrum to have a continuous spread with a width of ~ 70 G for nitroxides. Since the magnitude of the field jump in our TREPR spectrometer is restricted to ~ 30 G, the above procedure could not be performed.

To this end, we selected the on resonance position to be $x_1$ (fig. S4) and used a 15 G jump to the low field side as the off-resonance position, which safely corresponds to a non-resonance position (position $o_1$ in fig. S4). Thus, the true TREPR transient corresponds to

$$\text{True TREPR } transient = transient\ @x1 - transient@o1.$$

The only downside of position $x_1$ is that it corresponds to the low field side of the powder spectrum, where the intensity of the signal is smaller. Since the spectrum is most intense at position $x_2$, we repeated the same procedure at position $x_2$ too. In this case, we used our maximum jump value of 30 G (and moved to the high field side for the off-resonance transient) corresponding to point $o_2$. This resulted in a much better S/N signal, however, it has a downside that the position $o_2$ still has a EPR signal contribution and hence is not a clean off resonance position. For the present studies, we calculated the ESP enhancement at both positions $x_1$ and $x_2$ and report their average values.

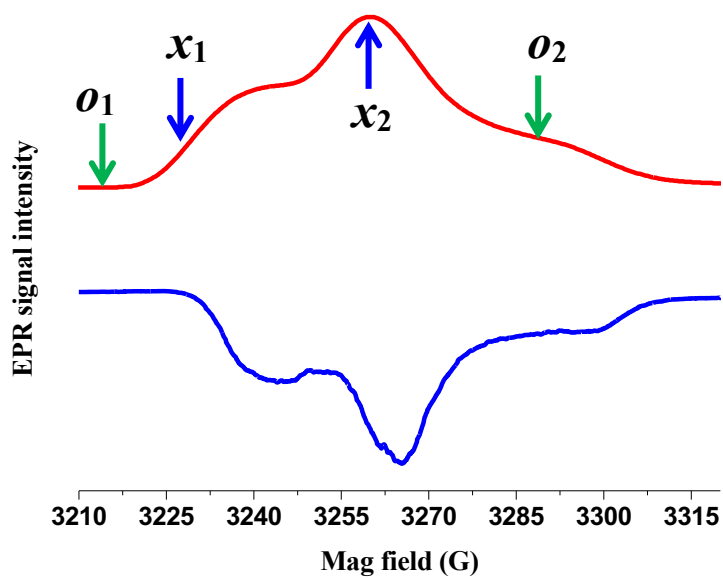

**Figure S4**: Experimental steady-state EPR spectrum (integrated form of 1st derivative) (red) and TREPR spectrum (blue) recorded 2.5 μs after the 355 nm laser pulse. Sample: ANCOOT in toluene at 100 K. The blue arrows denote $x_1$ and $x_2$ corresponding to the fixed magnetic field

values where the time profile of the electron hyperpolarization was recorded. The green arrows denote $o_1$ and $o_2$ corresponding to the fields that were used to perform the on-off subtraction procedure.

A typical TREPR plot of ANCOOT and the subsequent procedure to obtain the ESP enhancement is shown in Fig. S5.

Based on the $I_{pol}$ value for the excited state concentration ($S_1^0$) generated immediately after the laser pulse, the $I_{pol}^{MAX}$ for maximum concentration of excited state molecules is determined which occurs when $S_1^0 = R_{gnd}$ (where $R_{gnd}$ represents the ground state concentration). This value of $I_{pol}^{MAX}$ may be taken as a low bound estimate of $P_{Pol}/P_{eq}$. Carrying out the above procedure for the TREPR time profiles of ANCOOT in toluene at 100 K, the value of ESP enhancement, $P_{Pol}/P_{eq}$, were found to be comprised between 100 and 180.

Lastly, we wish to mention about the large deviation in the estimation of $P_{Pol}/P_{eq}$ value. One of the major sources of uncertainty was the laser energy reaching the sample. Some of the difficulties that compounded this problem were the need to limit laser pulses' energy to avoid saturating the microwave detector caused by the strong polarized signal. Additionally, the reduced pulse energy increased the shot to shot laser fluctuation and laser energy drifts), and the use of two different methods for laser energy measurement: rectangular slit and no slit with a geometrical correction factor (see experimental section) to account for the differences in the dimensions of the laser beam and the sample tube (The laser beam was circular with ~ 6.4 mm diameter while the sample tube was cylindrical with a 3 mm id and 6.4 mm length that is exposed to the laser (See fig. S4)). Note: there is additional losses due to scattering from the curved surfaces of the sample tube and the low temperature Dewar, which also adds to uncertainty of the measurements.

Thus, given the above sources of uncertainty, the difficulties associated with reproducing the Boltzmann signal for frozen sample and the fact that we used a qualitative method for estimating the $P_{Pol}/P_{eq}$ value, our lower bound estimate of the $P_{Pol}/P_{eq}$ value could only be considered as tentative. In future this value could be refined with hardware improvements such as an optical fiber guided laser irradiation directly to the top of the sample (to avoid diffraction losses at curved surfaces) and a rigorous kinetic Bloch formalism which considers the inhomogeneous broadening at low temperature.

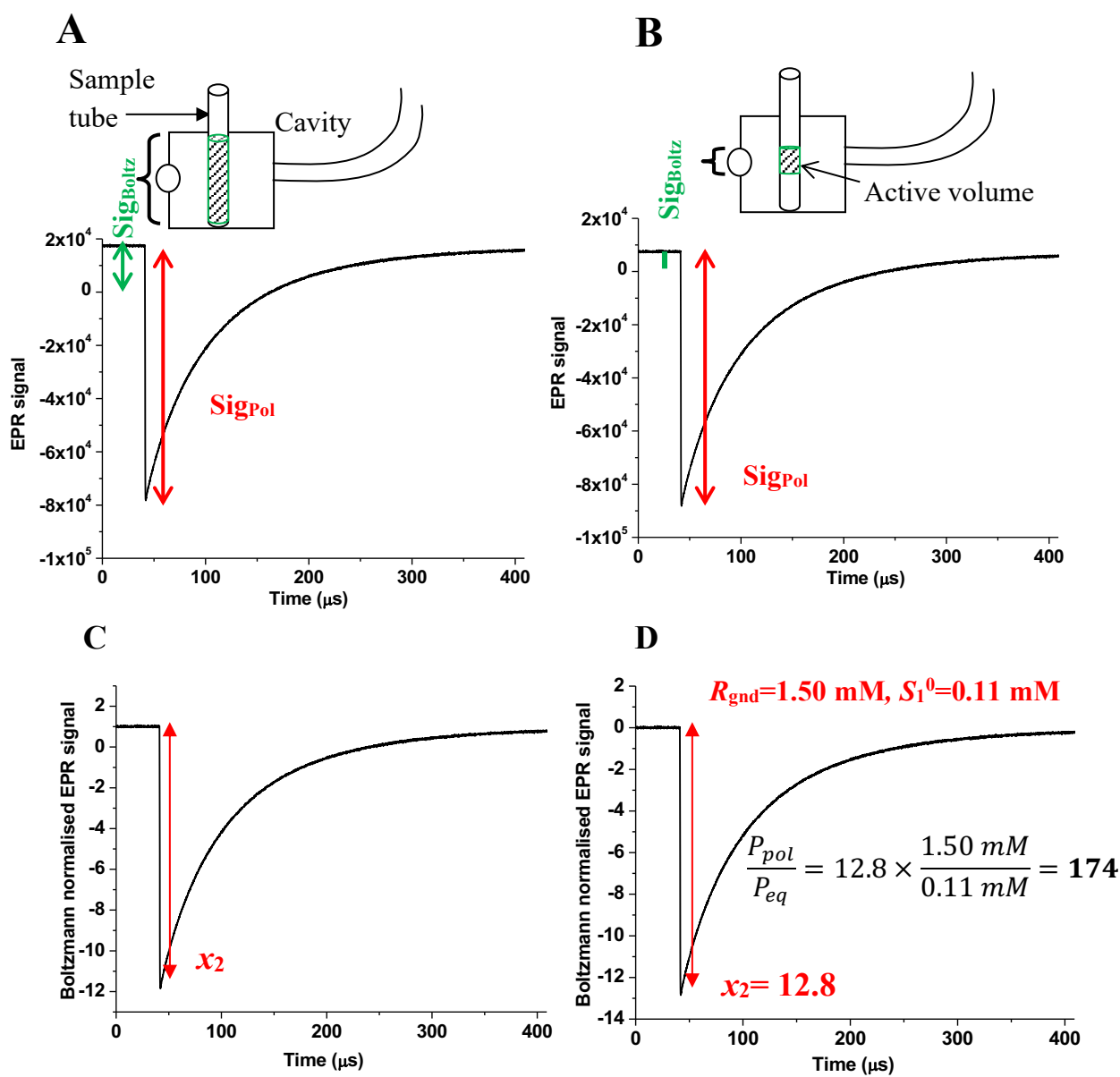

**Figure. S5.** Processing of the TREPR signal for obtaining a lower estimate of $P_{Pol}/P_{eq}$. A) As acquired time profile of the TREPR signal recorded at the position $x_2$ (fig. S4). Note the Boltzmann signal comes from the entire cavity while the polarized signal comes only from the active volume. B) The processed TREPR time profile, where the signal height before laser pulse corresponds to only those molecules which are capable of absorbing light i.e., within a circular cross section of 6.35 mm hole of the cavity. Approximately, this corresponds to 43% of the signal coming from the entire length of the sample within the cavity C) Plot of the normalized acquired signal with respect to Boltzmann signal. The height $x_2$ is used to obtain a lower estimate of $P_{Pol}/P_{eq}$. D) Plot of signal shifted down (minus one) such that the height of $x_2$ can be directly read from the $y$ axis. Concentrations: 1.5 mM (ground state), $S_1^0$=0.11 mM (immediately after laser excitation).

**Appendix 5: Materials and methods**

Steady-state and time-resolved EPR spectra were recorded in a laboratory-built X-band EPR spectrometer.[20] For the time-resolved EPR, the exciting source was the 3rd harmonic of an Nd:YAG laser (Quantel Model: YG-981C, wavelength: 355 nm, repetition rate: 15 Hz, energy at the sample: 1-4 mJ/pulse). Because of the extremely strong spin polarized EPR signals, the laser energy was reduced to its lowest output value. The absorbance of all the solutions at the exciting wavelength was less than 1. The samples were made in quartz tubes (O.D.: 4 mm, I.D.: ~ 3 mm, Wilmad Glass, USA) and degassed by three cycles of freeze-pump-thaw under a vacuum of $10^{-5}$ mbar. The low temperature experiments were performed by passing cold $N_2$ gas through a Dewar that contained the sample tube. The temperature at the sample tube was calibrated by inserting a thermocouple inside the cavity.

The EPR time profiles were collected after locking the magnetic field to the desired position, using a magnetic field-microwave frequency interlock (see appendix 3). Each transient was collected alternately at 'on-resonance' and 'off-resonance' by step modulating the magnetic field by 15-30 G at 15 Hz (fixed by the laser repetition rate). Normally, an AC-coupled amplifier blocks the steady-state EPR signal due to the Boltzmann population difference. However, the experiments, involving modulating the magnetic field synchronously with the laser repetition rate, was designed to ensure that the Boltzmann signal passed through the AC-coupled preamplifier and such that it could be recorded by a transient recorder. The true EPR signal was obtained by subtracting the 'off-resonance' curve from the 'on-resonance' curve. This subtraction also removes any non-resonant artefact that rides on the observed microwave signal, when the laser pulse enters the microwave cavity.

The TREPR time profiles were fitted by RQM kinetic model (See appendix 1). Most of the parameters needed for the simulation were either measured experimentally or obtained from literature. The spin-lattice relaxation times were measured by the pulse saturation-recovery technique. The microwave magnetic field $\omega_1$ for our cavity was found to be equal to $0.50\sqrt{P_w}$ for the cylindrical tube, where $P_w$ represents the microwave power (in mW) inside the cavity. The laser energy absorbed by the sample, which determines the initial excited state concentration, $S_1^0$, is a crucial parameter that determines the intensity of the ESP signal. Thus it was imperative to know the energy of the laser pulse reaching the sample inside the microwave cavity of the EPR spectrometer. It was obtained as follows:

**Calculation of laser energy incident on the sample**: We inserted a rectangular slit (I) of 6.35mm × 3.00 mm in front of the laser entry hole of the microwave cavity (Fig. S6). Here, 6.35 mm was the same as the diameter of the laser entry hole, and 3.00 mm was the I.D. of the EPR sample tube . The best alignment of the slit, the hole of the cavity and the EPR sample was achieved by maximizing the transient TREPR signal of ANCOOT. This ensured a collinear arrangement of the slit hole and the cavity hole, and the energy of the laser measured after the slit was taken to be the energy incident on the sample.

The rectangular slit matched the dimensions as the sample tube. However, its alignment needed a very critical positioning. Additionally, even after achieving the best alignment, we observed that the signal height reduced by ~1.4 times after insertion of the slit. Ideally, no reduction in signal should occure with a perfect alignment, as the dimensions of slit matched the sample tube. In view of this, the experiments were performed without the slit as to not sacrifice signal intensity. However, the laser energy measured without any slit was more than what the sample received, since the tube I.D. was only 3 mm (while the laser beam O.D was ~ 6.4 mm). So a geometrical correction factor, $X^{CF}$, was applied, see below.

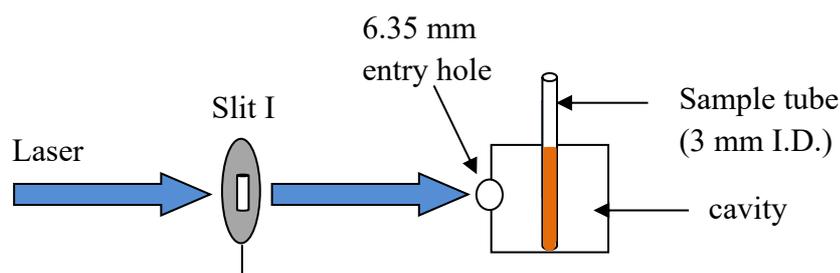

**Figure S6**: Arrangement of Slit in front of the microwave cavity for measurement of laser energy.

$X^{CF}$ represents the fractional energy of the laser beam that irradiated the sample (present in the 3.0 mm O.D. tube). To estimate $X^{CF}$, we first measured the laser energy after (i) a circular slit (II) of diameter (6.35 mm) and the rectangular slit (I) in a collinear geometry, and (ii) the rectangular Slit I only. The diameter of the circular slit was the same as that of the laser entry hole in the front surface of the cavity. This showed that about 64% of the energy after the circular slit passed through rectangular slit I. Since slit I resembles the tube dimensions, the correction factor was assumed to 0.64.

We found that the ESP enhancement obtained via the two sets of measurements differed by a factor our. The rectangular slit gave a lower estimate while the experiment without slit gave a larger estimate of the ESP enhancement. We do not understand the origin of this behaviour but it could be that the rectangular slit underestimates the $P_{Pol}/P_{eq}$ value (due to improper slit-tube alignment as pointed above) while the no slit case overestimates it. Thus, for the present work, we report the average of the two sets of measurements. In future, a more refined value could be obtained by use of an optical fiber guided laser irradiation directly on the top of the sample tube. This would give a better estimate of the laser energy reaching the sample. Note that this method would also overcome the losses of laser energy that are incurred due to diffraction at curved surfaces of the sample tube and the low temperature Dewar.

Further details about the processing of TREPR decay profiles are given in Appendix 3.

**Appendix 6: Nitroxide hyperpolarization mechanism**

Simulating the nitroxide hyperpolarization in the solid state is complex as the underlying details of the mechanism are unknown. Thus, a thorough model is beyond the scope of this article. Instead, we use an approximate model, that mimics the said hyperpolarization model.

To begin, we assume that the triplet state from the chromophore gets polarized at a given rate. This triplet state is in contact with the nitroxide and they tend to equilibrate as depicted in figure S7.

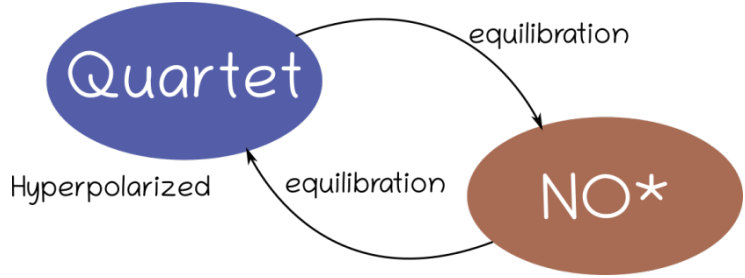

**Figure. S7**: Schematic of the hyperpolarization mechanism of the nitroxide. The hyperpolarized quartet state equilibrates its polarization with the nitroxide.

With such model, mimicking the optically induced hyperpolarization is relative straightforward. In the subspace spanned by the identity, the quartet polarization and the nitroxide polarization, $\{\hat{E}, \hat{Q}_z, \hat{S}_{z,a}^{NO}\}$, the evolution of the density matrix $\hat{\sigma}$ can be written as

$$\frac{\partial}{\partial t}\hat{\sigma} = \begin{pmatrix} 0 & 0 & 0 \\ T_z^0/T_{1,T} & -\frac{1}{T_{1,T}} - r & +r \\ P_{z,a}^0/T_{1,e}^{NO} & +r & \frac{1}{T_{1,e}^{NO}} - r \end{pmatrix} \hat{\sigma} \qquad \text{Eq. (S17)}$$

Where $r$ is the equilibration rate between the triplet and the nitroxide, $T_{1,T}$ is the rate at which the Triplet state is hyperpolarized, $T_z^0$ and $P_{z,a}^0$ are the hyperpolarized and equilibrium polarization of the triplet state under optical irradiation and the nitroxide respectively. In the case where $r$ is much greater than $\frac{1}{T_{1,e}^{NO}}$ at steady state we have

$$0 = \begin{pmatrix} 0 & 0 & 0 \\ T_z^0/T_{1,T} & -\frac{1}{T_{1,T}} - r & +r \\ 0 & +r & -r \end{pmatrix} \hat{\sigma} \qquad \text{Eq. (S18)}$$

which then leads to

$$\langle S_{z,a} \rangle = \text{trace}(\hat{\sigma}\hat{S}_{z,a}) \approx T_z^0. \qquad \text{Eq. (S19)}$$

This is experimentally supported by the significant nitroxide hyperpolarization that is observed experimentally in the solid-state. In addition, the simulations of $k_{DQ}{}^a$ for the appropriate exchange interaction $|J_{CR}|$ confirms that the polarization equilibration is extremely fast as compared to the nitroxide relaxation time $T_{1,e} \approx 0.3$ ms (see figure S7) even at high field. It is safe to assume that the nitroxide steady state polarization reaches a certain level under optical irradiation. In the actual MAS-DNP simulations, the triplet state can then be largely ignored and its effect on the nitroxide spin can be simply mimicked in a smaller subspace as $\{\hat{E}, \hat{S}_{z,a}^{NO}\}$ leading to:

$$\frac{\partial}{\partial t}\hat{\sigma}_{eff} = \begin{pmatrix} 0 & 0 \\ \frac{lT_z^0}{T_{1,eff}} & -\frac{1}{T_{1,eff}} \end{pmatrix}\hat{\sigma}_{eff} \qquad \text{Eq. (S20)}$$

Where $T_{1,eff}$ is an effective relaxation rate, and $l$ is a factor modulating the efficiency of the transfer of polarization from the Triplet state.

Therefore, the previously used MAS-DNP simulations can also be reused with minimal modifications. For the simulations we chose a conservative value for $T_{1,eff}$ and we set it to $1/k_{DQ}{}^a \sim 10$ μs.

This simplification enables the use of the previously developed Box-Model with a minor modification, and we only changed $lT_z^0$ when testing the different paramaters.

**Appendix 7: Box Model**

The simulations were carried out with a previously developed model. The details have been reported in two prior publications.[24,28] The basic building block of the Cross-Effect mechanism consists of a three-spin system: two electron spins and a nucleus. Such small spin system is sufficient to mimic the Cross-Effect induced by a biradical under MAS-DNP conditions.[21–24] However, this model does not consider the effect of the other biradicals in the vicinity. At typical biradical concentrations used for MAS-DNP (i.e. ca 10 mM), the presence of additional biradical modifies the spin dynamics of the three spin system and the existence of an extended dipolar network affects the spatial distribution of the electron spin polarization.

To simulate the impact of those biradical a model containing many copies of the 3-spin system is used. In this box each of the three-spin system has its own crystal orientation and can be interacting with the others. Such "box" containing 40 biradicals is represented in figure S8.

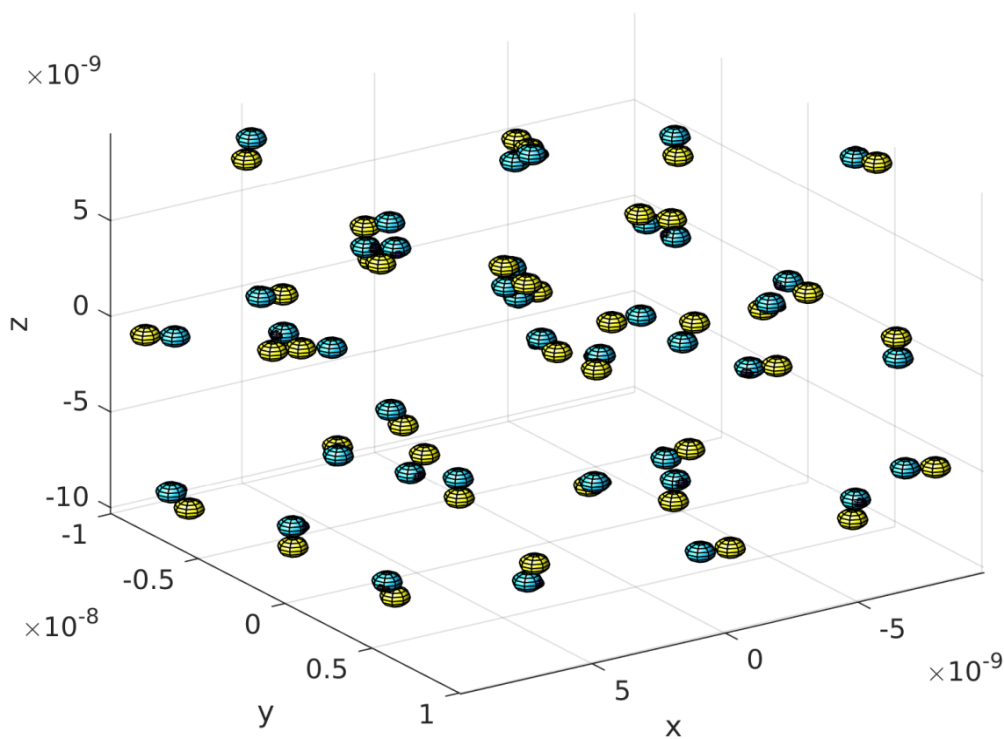

**Figure. S8**: Example of a box containing 40 copies of the three-spin system (2 electrons (blue and yellow) spin – 1 nucleus (tiny red sphere)).

To avoid using very large Liouville space basis, the Landau-Zener approximation is used. Additional details about the box model can be found in ref. [24]

**Appendix 8: Numerical Simulation parameters and definitions**

Experimental conditions:

- Temperature 100 K,
- microwave frequency 526.9 GHz,
- MAS frequency 8 kHz.
- Except for MAS-DNP field profiles, the magnetic field for the simulations was set to 18.8 T.
- In cases where microwave irradiation were considered the nutation frequency was set to 0.2 MHz.

Hetero-biradical case:

- *g* tensors: [2.0095,2.0061,2.0021] for the nitroxide and [2.0032,2.0030,2.0027] for the Trityl moiety.
- *g*-tensor relative orientation $[\alpha, \beta, \gamma]$ = [90 90 90] degrees.
- Dipolar coupling nitroxide-trityl: 30 MHz[29].
- Minimal distance between radical pair from two different biradicals: 4.2 nm.
- Dipolar angles $[\theta, \phi]$ = [90 180].
- The nitroxide was assumed to be connected to a proton with a dipolar hyperfine coupling of 3 MHz.
- Hyperfine dipolar vector oriented along $[\theta_n, \phi_n]$ = [90 0].
- $T_{1,e}$ relaxation times was set to 1 ms for Trityl.
- $T_{1,eff}$ relaxation times was set to 10 µs for the nitroxide
- $T_{2,e}$ relaxation times of was set to 2.5 µs for both radicals.
- Proton relaxation times was assumed to be $T_{1,n} = 0.1$ s.

AMUPol

- *g* tensors : [2.00923,2.00619,2.00212].
- *g*-tensor relative orientation $[\alpha, \beta, \gamma]$ = [58 57 126] degrees.
- Dipolar coupling: 35 MHz.
- the dipolar angles $[\theta, \phi]$ = [78 167].
- Exchange interaction: $J_{a,b} = -16$ MHz.
- The nitroxide *a* was assumed to be connected to a proton with a dipolar hyperfine coupling of 3 MHz
- Hyperfine dipolar vector oriented along $[\theta_n, \phi_n]$ = [0 90]
- $T_{1,e} = 0.3$ ms for the second nitroxide.
- $T_{1,eff}$ relaxation times of was set to 10 µs nitroxide *a*.
- $T_{2,e}$ relaxation times was set to 2.5 µs for both radicals.
- Proton relaxation times was assumed to be $T_{1,n} = 0.1$ s.

**Appendix 9: Derivation of the field dependence of $|\epsilon_B|$**

With the chosen MAS-DNP simulation parameters, the nuclear polarization at the quasi-periodic steady state follows the following equation[23]

$$|P_{e,a} - P_{e,b}| = |P_n| \qquad \text{Eq. (S21)}$$

Where $|P_a - P_b|$ is the maximum polarization difference between the two electron spins across one rotor period at steady state.[23]

If we assume that we have electron $a$ that tends to a hyperpolarized level $P_{e,a} \to P_{\text{hp}}$, while $b$ remains close to the thermal equilibrium, then the enhancement is given by:

$$|\epsilon_B| = \frac{|P_n|}{|P_n^{\text{eq}}|} \to \frac{|P_{\text{hp}} - P_{e,b}^{\text{eq}}|}{|P_n^{\text{eq}}|}. \qquad \text{Eq. (S22)}$$

The polarization at equilibrium depends on the external magnetic field strength, and the temperature and it is defined as:

$$|\epsilon_B(T)| = \frac{|B_a^{\text{EHP}}(T) - B_0|}{|B_0|} \qquad \text{Eq. (S23)}$$

in which $B_a^{\text{EHP}}$ is an effective field for a given temperature $T$. Since the dynamics of the spin system, in particular the dipolar coupling in between the biradicals, affects $|P_{e,a} - P_{e,b}|$,[23,25] we can simply assume that a scaling factor must be accounted for, hence

$$|\epsilon_B(T)| = \frac{|B_{a,\text{eff}}^{\text{EHP}}(T) - B_0|}{|B_0|}. \qquad \text{Eq. (S24)}$$

This equation was used to fit the data in figure 3B.

**Appendix 10: Additional results**

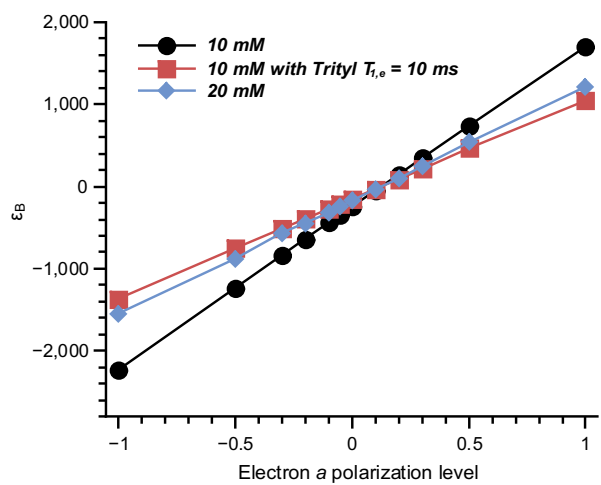

**Figure. S9**: Simulations of the MAS-DNP polarization gain $\epsilon_B$ for Trityl-TEMPO using the modified Box model. Evolution of the maximum $\epsilon_B$ as a function of the nitroxide polarization level for 10 mM (black circle), 20 mM (blue diamond) and 10 mM with longer Trityl's relaxation time T1e = 10 ms (red square).

Figure S9 report the effect of the Trityl's relaxation time or the effect of the biradical's concentration on $\epsilon_B$. The longer electron relaxation and/or the higher biradical concentration generates more smearing of the electron spin polarization difference as previously predicted,[25] generating lower $\epsilon_B$.

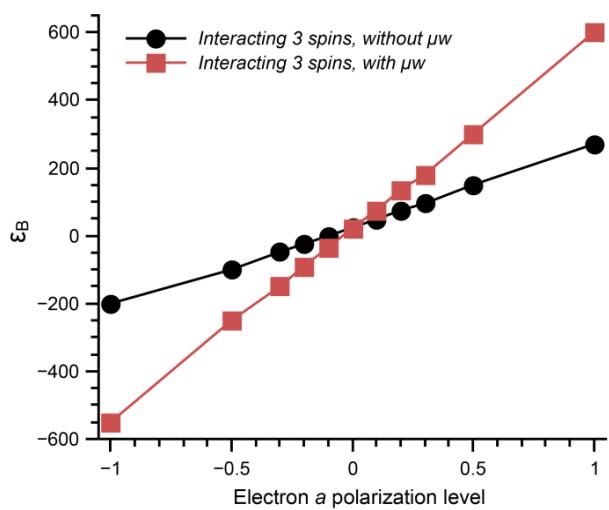

**Figure. S10**: Simulations of the MAS-DNP polarization gain $\epsilon_B$ for AMUPol-like, using the modified Box-Model. Evolution of the maximum $\epsilon_B$ in the case of the interacting 3-spin model case without μw (full black circles), and with μw irradiation transition (full red squares) as a function of the nitroxide polarization level.

Figure S10 reports the effect of the electron spin hyperpolarization level on the nuclear polarization using AMUPol's previously described structure.[26] This result is remarkable. AMUPol is known to generate significant nuclear depolarization, which means that it is not the best biradical to generate large electron polarization differences[27,28]. One would expect that the nitroxide reaches some steady state polarization where both moieties would have similar polarization levels. Surprisingly, the hyperpolarization is not entirely redistributed thus NHP is observed. The simulations revealed that the "inter-biradical" interaction had minimal effect on the NHP predictions which shows that most of the polarization difference between the nitroxides is lost during the "internal" biradical spin dynamics. This is understandable as AMUPol is known to generate significant nuclear depolarization in absence of microwave irradiation, as both electrons' spins tend to have moderate electron polarization difference at steady state.[27,28] This is due in part to the moderate distance between the g-tensors[29]. From this observation, one may expect major complications nonetheless due to potential effect of relative orientation or the strength of the exchange interaction. These questions remain beyond the scope of this work.